\documentclass[review,12pt]{elsarticle}

\usepackage{graphicx}
\usepackage{multirow}
\usepackage{multicol}
\usepackage{booktabs}
\usepackage{color}

\usepackage[british]{babel}
\usepackage[colorlinks=true,linkcolor=black, citecolor=blue, urlcolor=blue]{hyperref}

\makeatletter
\def\ps@pprintTitle{%
 \let\@oddhead\@empty
 \let\@evenhead\@empty
 \def\@oddfoot{\centerline{\thepage}}%
 \let\@evenfoot\@oddfoot}
\makeatother

\begin{document}

\begin{frontmatter}

\title{Comparing reverse complementary genomic words
 based on their distance distributions and frequencies\\ \vspace{1cm}}

\author{Ana Helena Tavares\fnref{mat}\corref{cor1}}
\author{Jakob Raymaekers\fnref{ku}}
\author{Peter Rousseeuw\fnref{ku}}
\author{Raquel M. Silva\fnref{ibi}\fnref{iee}}
\author{\\Carlos A.C. Bastos\fnref{iee}\fnref{det}}
\author{Armando Pinho\fnref{iee}\fnref{det}}
\author{Paula Brito\fnref{up}}
\author{Vera Afreixo\fnref{mat}\fnref{ibi}\fnref{iee}}

\fntext[mat]{Department of Mathematics \& CIDMA, University of Aveiro, Portugal}
\fntext[ku]{Department of Mathematics, KU Leuven, Belgium}
\fntext[ibi]{Department of Medical Sciences \& iBiMED, University of Aveiro,Portugal}
\fntext[iee]{Institute of Electronics and Informatics Engineering of Aveiro (IEETA)}
\fntext[det]{Department of Electronics, Telecommunications and Informatics, University of Aveiro, Portugal}
\fntext[up]{Faculty of Economics \& LIAAD - INESC TEC, University of Porto,Portugal}
\cortext[cor1]{Corresponding author}


\begin{abstract}
In this work we study reverse complementary genomic word pairs in the human DNA, by
comparing both the distance distribution and the frequency of a word to those of its
reverse complement. Several measures of dissimilarity between distance distributions
are considered, and it is found that the peak dissimilarity works best in this setting.
We report the existence of reverse complementary
word pairs with very dissimilar distance
distributions, as well as word pairs with very similar distance distributions even
when both distributions are irregular and contain strong peaks. The association
between distribution dissimilarity and frequency discrepancy is explored also, and it is
speculated that symmetric pairs combining low and high values of each measure may
uncover features of interest. Taken together, our results suggest that some
asymmetries in the human genome go far beyond Chargaff's rules. This study uses both
the complete human genome and its repeat-masked version.
\end{abstract}


\end{frontmatter}

\section{Introduction}\label{intro}
The analysis of DNA sequences is an extremely broad
research domain which has seen several new approaches
over the last years. One of these newer approaches is
the study of distance distributions of genomic words.
A genomic word, also called an oligonucleotide, is a
sequence of nucleotides which are represented by the
letters $\{A,C,G,T\}$.
In DNA segments, the inter-word distance is defined as
the number of nucleotides between the first symbol of
consecutive occurrences of that
word~\cite{afreixo2009,Tavares2015}.
For instance, in the DNA segment
$A\underline{CG}T\underline{CG}ATC\underline{CG}TG
 \underline{CG}\,\underline{CG}$ the
inter-$CG$ distances are (3,5,4,2).
For each word, all of its inter-word distances in the genome sequence
can be counted and aggregated into a \emph{distance distribution},
which contains the frequency of each distance.
These distributions provide a characterization of
genomic words which can be studied using statistical
techniques for probability density functions.\\

In this paper we are particularly interested in
the study of {\it symmetric word pairs}.
A symmetric word pair is formed by a word $w$
and its reverse complement $\bar{w}$, which is
the word obtained by reversing the order of the
letters and interchanging the complementary
nucleotides $A \leftrightarrow T$
and $C \leftrightarrow G$. For instance, the
reverse complement of $w=AAGT$
is $\bar{w}=ACTT$, and together they form the
symmetric pair $\{w,\bar{w}\}$.
The interest in these pairs stems from Chargaff's
second parity rule which implies that within a
strand of DNA the number of complementary
nucleotides is similar \cite{forsdyke2000}.
One potential explanation postulates that this
phenomenon would be an original feature of the
primordial genome, the most primitive nucleic acid
genome, and the preservation of strand symmetry
would rely on evolutionary mechanisms
\cite{zhang2010strand}.
Symmetric word pairs can occur in a genome through recombination events such as
duplications, inversions and inverted transpositions~\cite{baisnee2002,albrecht2007}.
These segments have been associated with specific biological functions, namely,
replication and transcription, and major evolutionary events including recombination and
translocations. Also, the potential to form secondary DNA structures can cause the
genome instability observed in some diseases~\cite{inagaki2016}.

Chargaff's second parity rule has led to the natural
question whether this also holds for symmetric
word pairs.
This question has been answered to a certain extent
in the existing literature~\cite{afreixo2013breakdown,afreixo2015analysis,albrecht2006,baisnee2002}, as it has been observed
that even for long DNA words in several organisms,
including the human genome, the frequency of a
word is typically (but not always) similar to
that of its reverse complement.
However, two words with the same frequency in a
sequence may exhibit very distinct distance
distributions along that sequence. This leads to
the natural follow-up question: do symmetric word
pairs have similar distance distributions?\\

Tavares et al.~\cite{Tavares2015} addressed
this question for words of length $k \leq 5$
in the human genome. Adopting a whole-genome
analysis approach, the discrepancy between distance
distributions was evaluated using an effect
size measure. The authors concluded that the
dissimilarity between the distributions of symmetric word pairs of
this length was negligible. The authors also
reported that for each word $w$, the distance
distribution nearest to the distance distribution of
$w$ is most often that of $\bar{w}$, the
reverse complement of $w$.\\

\begin{figure}[htbp]
\centering
  \includegraphics[width=0.50\textwidth]{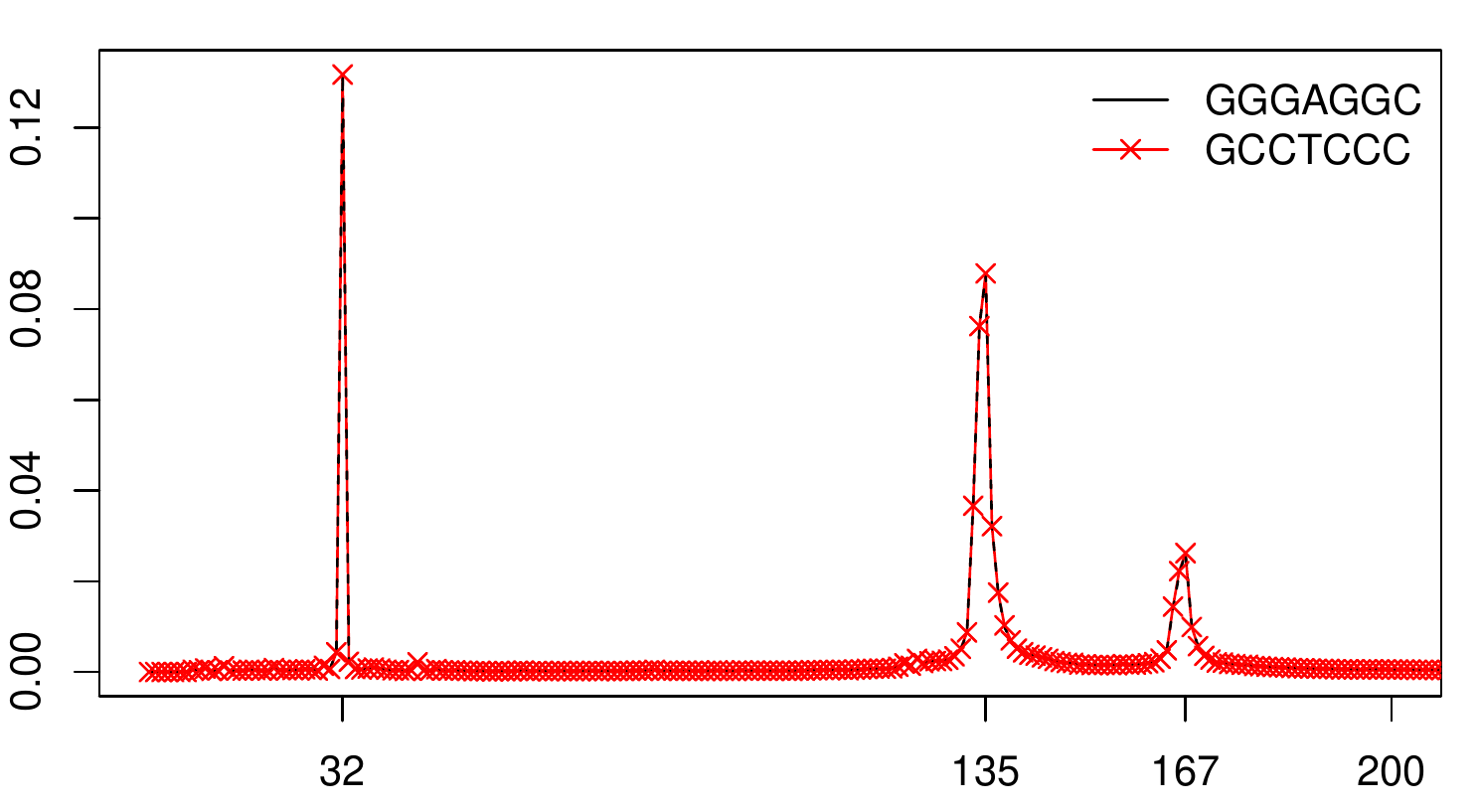}	
\caption{Distance distribution of the genomic word
  $w=GGGAGGC$ and of its reverse complement
	$\bar{w}=GCCTCCC$ in the human genome (adapted from~\cite{Tavares2015}).}
\label{fig:example}
\end{figure}

As an example, Figure~\ref{fig:example} shows
the distance distribution of the word
$w=GGGAGGC$ in the human genome.
Its peaks correspond to three distances that
occur much more often than others.
In this example the distance distribution of the
reverse complement $\bar{w}=GCCTCCC$ is
extremely similar.

In order to study differences between distance
distributions, a new dissimilarity measure was
proposed by Tavares et al.~\cite{tavares2017pacbb}. Based on
the gaps between the locations of their peaks
and the difference between the sizes of these
peaks, the {\it peak dissimilarity} becomes
high when the distributions have very different
peaks, or when one distribution has strong
peaks and the other does not.
In this article we extend their work in two ways.
First, we compare the peak dissimilarity with
two earlier dissimilarity measures and argue
for its superiority in the analysis of distance
distributions between symmetric word pairs.
Secondly, we combine the peak dissimilarity with
information about the frequencies of the word
and its reverse complement to improve the
identification of atypical genomic word pairs.
We also draw a comparison between the observed
distribution and the expected distribution
under randomness. Using these techniques we
detect several atypical word pairs, which
we annotate by identifying the chromosomes
and genes where their differences are most
pronounced.\\

The paper is organized as follows. In Section
2 we describe measures of the discrepancy
between frequencies and distance distributions,
including the peak dissimilarity. Section 3
compares the behavior of these dissimilarity
measures in our particular research problem.
Section 4 identifies and investigates the
symmetric word pairs that are most and least
dissimilar, using both their frequencies and
their distance distributions. It also explores how
well the results hold up in a masked sequence.
Section 5 concludes.

\section{Measures of dissimilarity}
\subsection{Discrepancy between word frequencies}
To measure the discrepancy between the total
absolute frequencies of reverse complementary
words $w$ and $\bar{w}$, we count all
occurrences of each word along the DNA sequence.
The number of times $w$ occurs is denoted as
$n^w$, and that of $\bar{w}$ is $n^{\bar{w}}$.
Under the null hypothesis that the true
underlying probabilities of $w$ and $\bar{w}$
are equal, the expected frequency of $w$ is
$e = (n^w+n^{\bar{w}})/2\;$.
The Pearson residual~\cite{agresti2007} of $w$ is then given by
$(n^w - e)/\sqrt{e}\;$. The {\it absolute
Pearson residual} (APR) of $w$ is thus
\begin{equation}
\mbox{APR}(w)= \frac{|n^w - e|}{\sqrt{e}}
 = \frac{|n^w-n^{\bar{w}}|}
         {\sqrt{2(n^w+n^{\bar{w}})}} \;\;.
\end{equation}
Note that $\mbox{APR}(w)=\mbox{APR}(\bar{w})$
and that $2 \mbox{APR}^2(w)$ equals the usual
chi-squared statistic for testing the equality
of the underlying probabilities.

\subsection{Dissimilarity measures for distance
            distributions}
Assuming that the DNA sequence is read through
a sliding window of word length $k$, the
inter-word distance sequence is defined as the differences between
the positions of the first symbol of consecutive
occurrences of that word.
For instance, the inter-$CG$ distances sequence in the
DNA segment $CGTACGCGACG$ is (4,2,3).
The distance distribution of $w$, denoted by $f^w$, gives
the relative frequency of each distance, i.e. the number
of times a certain distance occurs divided by the total
number of occurrences of the word $w$.

The word structure influences the distance distribution,
as some distances from 1 to $k$ may be absent. As an
example, note that the inter-$AAA$ distance can be equal
to one, but cannot be two or three. So, for words of
length $k$ we will only consider distances greater
than $k$.

We now wish to compare the distance distribution of each
word $w$ with the distance distribution of $\bar{w}$.
For this we describe three dissimilarity measures,
two of which have been used for a long time and one
is new.

\subsubsection{Euclidean distance}
The Euclidean distance is a standard tool which is
also used between distributions.
In our situation, the discrete probability
distributions $f^w$ and $f^{\bar{w}}$ have the same
domain. The word `discrete' refers to the domain,
as the distances are always integers.
The probabilities (i.e. frequencies) of a distance $i$
are denoted as $p_i = f^w(i)$
and $q_i =f^{\bar{w}}(i)$.
Then the Euclidean distance $D_E(f^w,f^{\bar{w}})$
is obtained by summing the squares of the
frequency differences:
\begin{equation}
D_E(f^w,f^{\bar{w}})=\sqrt{\sum_i (p_i-q_i)^2}
\label{eq:DE}
\end{equation}

\subsubsection{Jeffreys divergence}
The Kullback-Leibler divergence~\cite{kullback1951}
between $f^w$ and $f^{\bar{w}}$ is given by
$$D_{KL}(f^w,f^{\bar{w}})
  =\sum_i p_i \log(p_i/q_i)$$
where the $0\log0 =0$ convention is adopted.
The Kullback-Leibler divergence stems from
information theory.
It is always nonnegative and becomes zero when the
distributions are equal, and it is widely used as a
divergence measure between distributions.
But it is not symmetric, as
$D_{KL}(f^w,f^{\bar{w}})$ need not equal
$D_{KL}(f^{\bar{w}},f^w)$.
Therefore we will use a symmetrized version called
the Jeffreys divergence~\cite{jeffreys1946}:
\begin{equation}
D_J(f^w,f^{\bar{w}})=D_{KL}(f^w,f^{\bar{w}})+
          D_{KL}(f^{\bar{w}},f^w)\;\;.
\label{eq:DJ}
\end{equation}
Note that $D_J$ is not well defined if some
$p_i$ or $q_i$ are zero. In practice this can be
avoided by replacing the zero values by a small
positive value.
The Jeffreys divergence $D_J$ is a semimetric,
meaning that it
is symmetric, nonnegative, and reduces to zero
when the two distributions are identical.

\subsubsection{Peak dissimilarity}
The distance distributions $f^w$ and $f^{\bar{w}}$
may present several peaks, i.e., distances with
frequencies much higher than the global tendency
of the distribution, as we saw in
Fig.~\ref{fig:example}.
To describe the recently proposed peak
dissimilarity~\cite{tavares2017pacbb} we go
through three steps.\\

{\it 1. Identifying peaks.}
To determine peaks we slide a window of fixed
width $h$ along the domain of the distribution.
In each such interval of width $h$ we average
the absolute values of the differences between
successive frequencies, and call the result the
{\it size} of the peak on that interval.
The peak's {\it location} is defined as the midpoint of the interval. The strongest peak
is then determined by the interval with the highest size. For the second strongest peak
we only consider intervals that do not overlap with
the first one, and so on.

The bandwidth $h$ is a tuning parameter which
controls the number of
consecutive frequencies that are aggregated in
a region. There
is no best bandwidth, and different bandwidths
can reveal different features of the data. To
illustrate the effect of $h$ on peak
identification, consider the distance
distribution of the word $w=GGGAGGC$
in Figure~\ref{fig:example} which has a local
maximum
at distance 135.
When $h \leq 3$ the region around distance 135
gives rise to two intervals with high
peak size. However, when $h \geq 4$ these high
frequencies are combined into a single peak.
\\

{\it 2. Dissimilarity between two peaks.}
To measure the dissimilarity between two peaks
we take into account the difference between their
sizes and between their locations. Consider the
distance distributions $f^w$ and $f^{\bar{w}}$ which
are defined on the same domain with length $R$.
Let $t^w_i$ be a peak of $f^w$ with location
$l_i$ and size $v_i$ and let $t^{\bar{w}}_j$
be a peak of $f^{\bar{w}}$ with location
$\bar{l_j}$ and size $\bar{v_j}\;$.
To measure the dissimilarity
between these peaks we propose to use
\begin{equation}
  d(t^w_i, t^{\bar{w}}_j)=
	\left(\frac{|l_i-\bar{l_j}|}{R}+1\right)
	\left(\frac{|v_i-\bar{v_j}|}
	{\min\{v,\bar{v}\}}+1\right)-1
\label{eq:distD}
\end{equation}
where $v$  and $\bar{v}$ are the highest peak
sizes observed in each distribution.
If the peaks have the same location the
dissimilarity is reduced to a relative size
difference
$|v_i-\bar{v_j}|/\min\{v,\bar{v}\}$, and if
they have the same size it is reduced to a
relative location
difference $|l_i-\bar{l_j}|/R$. The denominator
$\min\{v,\bar{v}\}$ yields a high dissimilarity
when one distribution has strong peaks
and the other doesn't.\\

{\it 3. Peak dissimilarity between two
distributions.}
To measure the dissimilarity between two
distributions we compare their $n$ strongest
peaks, for fixed $n$. We propose
\begin{equation}
  D_P(f^w, f^{\bar{w}})=
	  \min_{\pi\in\mathcal{P}_n}
    \{\,\sum_{i=1}^n d(t^w_i,
		t^{\bar{w}}_{\pi(i)}) \,\}
\label{eq:d}
\end{equation}
where $\pi$ is a permutation of the indices
$i=1,\ldots,n$ meaning that $\pi(i)$ is the
image of $i$. The minimum is taken over the set
$\mathcal{P}_n$ of all permutations $\pi$
of $n$ elements.
In Fig.~\ref{fig:example} the minimum
in~(\ref{eq:d}) is attained for the simple
permutation $\pi(1)=1$, $\pi(2)=2$,
$\pi(3)=3$ yielding a tiny dissimilarity.
In general the proposed
measure~(\ref{eq:d}) depends on $n$,
the number of peaks considered, and on the
bandwidth $h$ used in the peak search.
Like $D_J$ also $D_P$ is a semimetric,
which is why we call it a
`dissimilarity' rather than a `distance'.

\subsection{Data and data preprocessing}
In this study we used the complete genome assembly,
build GRCh38.p2, downloaded from the website of the
National Center for Biotechnology Information\\
(http://www.ncbi.nlm.nih.gov/genome). We also used
pre-masked data available from the UCSG Genome
Browser (http://genome.ucsc.edu), in which the
repeats determined by Repeat
Masker~\cite{repeatMasker} and Tandem Repeats
Finder~\cite{benson1999} were replaced by N's.

The chromosomes were processed as separate sequences
and non-ACGT symbols were used as sequence separators.
The counts of word distances were generated using the C
language, taking overlap between successive words
into account and setting the maximal distance to 1000.
The R language was used to compute the distance
distributions, the dissimilarity measures and to
perform the statistical analysis.

\section{Comparison of dissimilarity measures}
In this section we will compare the dissimilarity
measures of Section 2 on the data under study,
consisting of all words of lengths 5, 6, and 7
in the human genome. In particular, the peak
dissimilarity is computed with bandwidth $h=5$
which revealed the essential peak structure
of the data, by capturing both ``isolated'' and
``grouped'' high frequencies.
The results are not overly sensitive to this choice,
and in fact very similar results were obtained for
$h=4,5,6$.
Also, we used the $n=3$ strongest peaks (for
$n=4,\ldots,7$ we obtained similar results in
much higher computation time).

\subsection{Correlation analysis}
For every symmetric word pair $\{w,\bar{w}\}$,
each of the four dissimilarity measures
provides a value.
These are the frequency discrepancy $APR$,
Euclidean distance $D_E$,
Jeffreys divergence $D_J$, and
peak dissimilarity $D_P$.
To evaluate the agreement between these four
measures we compute Spearman's rank
correlation coefficient $r_S$ between each pair.
For instance, to compare $APR$ and $D_E$ we rank
the values of each of them, and then compute
the product-moment correlation between these
two vectors of ranks.
Comparing each pair of measures yields the
Spearman correlation
matrices in Table~\ref{tab:spearman_complete},
one for each word length $k=5,6,7$.

\begin{table}[htbp]
\centering
\scriptsize
\caption{Spearman rank correlation
	matrices for frequency discrepancy $APR$ and distance
	distribution dissimilarities $D_E$, $D_J$,
	and $D_P$, by word length.}
\renewcommand{\arraystretch}{1.2}
\setlength\tabcolsep{2.5pt}
\begin{tabular}{|c|cccc|c|cccc|c|cccc|}
\hline
\multicolumn{5}{|c|}{k=5}  &
    \multicolumn{5}{|c|}{k=6} &
		\multicolumn{5}{|c|}{k=7} \\
\cline{2-5}\cline{7-10}\cline{12-15}
  & $APR$ & $D_E$ & $D_J$ & $D_P$ &
  & $APR$ & $D_E$ & $D_J$ & $D_P$ &	
  & $APR$ & $D_E$ & $D_J$ & $D_P$ \\	
\hline
  $APR$ & 1 &  &  &  &
  $APR$ & 1 &  &  &  &
  $APR$ & 1 &  &  &  \\
  $D_E$ & 0.635 & 1 &  &  &
	$D_E$ & 0.551 & 1 &  &  &
	$D_E$ & 0.283 & 1 &  &  \\
  $D_J$ & 0.573 & \bf{0.988} & 1 &  &
	$D_J$ &	0.403 & \bf{0.962} & 1 &  &
	$D_J$ & 0.029 & \bf{0.904} & 1 &  \\
  $D_P$ & 0.663 & 0.836 & 0.800 & 1 &
	$D_P$ & 0.622 & 0.784 & 0.678 & 1 &
	$D_P$	& 0.457 & 0.641 & \bf{0.427} & 1 \\
\hline
\end{tabular}
\label{tab:spearman_complete}
\end{table}

Overall the correlations decrease with increasing
word length, with $D_E$ and $D_J$ remaining the
most correlated ($r_S>0.90$).
The rather high correlation between $D_E$ and
$D_J$ may perhaps be explained by the formal
analogy between $D_E^2 = \sum_i (p_i-q_i)^2$
and  $D_J = \sum_i (p_i-q_i)(\log p_i-\log q_i)$.
By comparison $D_P$ is less correlated with
either of them, especially for $k=7$.
The correlation between $APR$ and the measures
$D_E$, $D_J$ and $D_P$ lies in between.
We may conclude that the various measures yield
complementary information, with the possible
exception of $D_E$ and $D_J$.
Therefore the adopted measure(s) should take
into account the features that are considered
important for the subject matter.
In the next subsection we will argue which
dissimilarity measures are the most useful in
the context of the present research problem.

\subsection{Comparing top-ranked sets}
For each distance distribution dissimilarity
measure ($D_E$, $D_J$ and $D_P$) we now
rank the dissimilarity values from smallest
to largest.
The highest ranks correspond to the most
dissimilar word pairs for that particular
dissimilarity measure. For instance, the
top 10\% ranked set for $D_E$ consists of
the word pairs whose Euclidean distance
exceeds the 90th percentile of $D_E$.
As discussed earlier, the ranks of $D_E$ and
$D_J$ are more correlated than those of $D_P$
and $D_J$ (see
Table~\ref{tab:spearman_complete}).
One way to assess whether the most dissimilar
distributions are the same in each top-ranked
set (regardless of their position within
that set) is to count the number of common
word pairs in those sets.
In particular, Table~\ref{tab:ranks} records
the fraction of common elements in the top
1\% ranked sets for $D_E$ and $D_J$ (under
the heading $R_{E,J})$, etc.
The top 1\% ranked sets for $D_E$ and $D_J$
indeed have the largest overlap, whereas
those of $D_J$ and $D_P$ have the least in
common, especially for $k = 6$ and $k = 7$.
The results for the top 10\% ranked sets are
similar.

\begin{table}[htbp]
\centering
\scriptsize
\caption{Comparison between the
rankings for $D_E$, $D_J$ and $D_P$: fraction
of common elements in the top 1\% and top 10\%
ranked sets.}
\renewcommand{\arraystretch}{1.2}
\setlength\tabcolsep{2.5pt}
\begin{tabular}{|c|ccc|ccc|ccc|}
\hline
  \multicolumn{1}{|c|}{} &
	\multicolumn{6}{|c|}{Overlap in top-ranked sets} \\
\cline{2-7}	
  \multicolumn{1}{|c|}{} &
	\multicolumn{3}{|c|}{top 1\%} &
	\multicolumn{3}{|c|}{top 10\%}  \\
\hline
  $\;\;k\;\;$ &
	$R_{E,J}$ & $R_{E,P}$ & $R_{J,P}$ &
  $R_{E,J}$ & $R_{E,P}$ & $R_{J,P}$ \\
\hline
  5 & 0.98 & 0.49 & 0.49 & 0.89 & 0.66 & 0.63 \\
  6 & 0.12 & 0.24 & \bf{0.05} & 0.63 & 0.61 & \bf{0.38}  \\
  7 & 0.23 & 0.03 & \bf{0.00} & 0.58 & 0.47 & \bf{0.18}  \\
\hline
\end{tabular}
\label{tab:ranks}
\end{table}

\begin{figure}[htbp]
\centering
\begin{tabular}{ccc}
  \includegraphics[width=0.32\textwidth
  ,trim={0.5cm 0.5cm 0.5cm 0.4cm},clip]
	{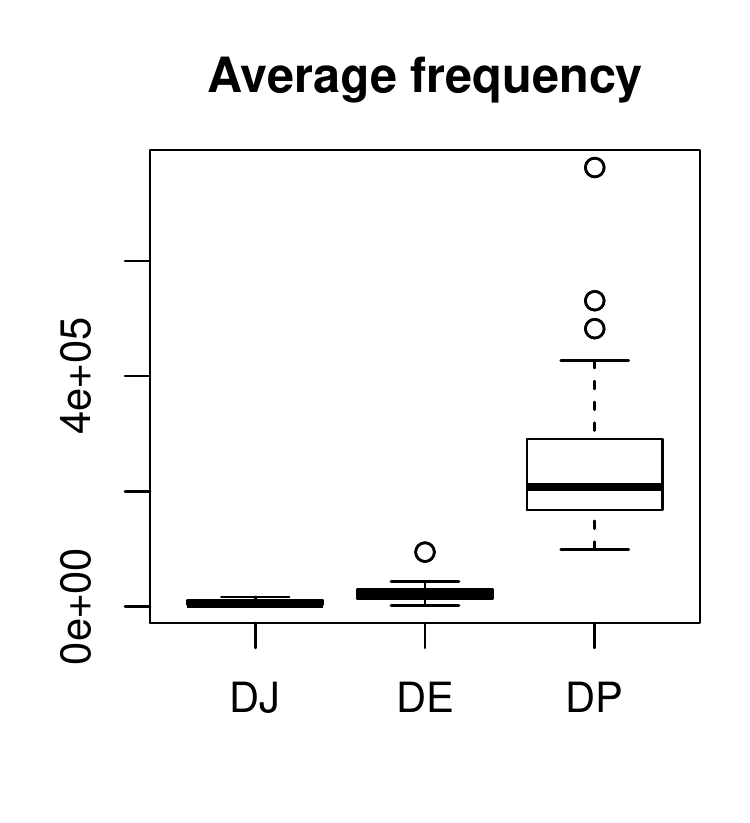} & &
  \includegraphics[width=0.32\textwidth
  ,trim={0.5cm 0.5cm 0.5cm 0.4cm},clip]
	{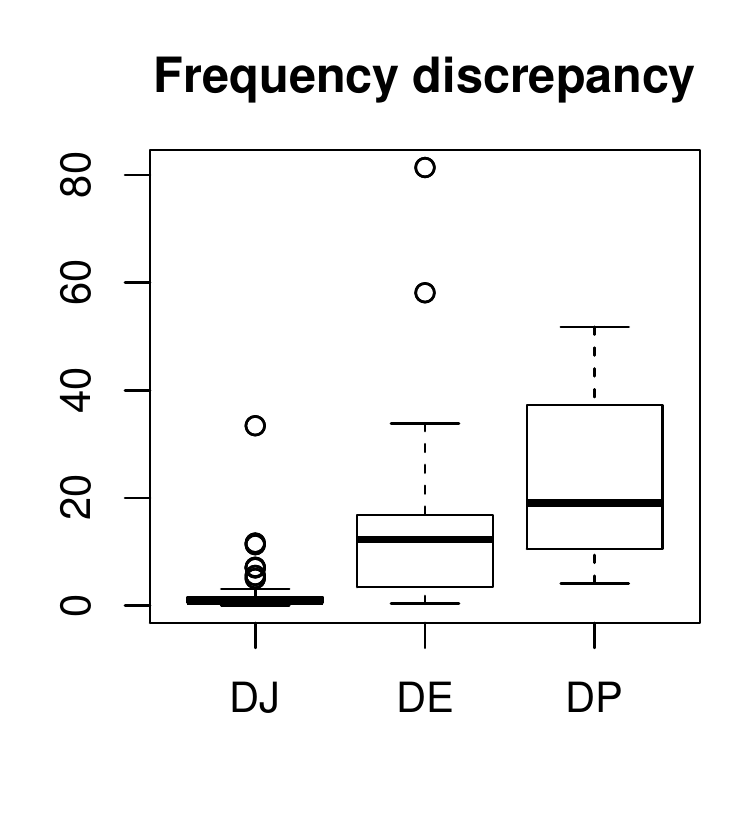}\\
 (a) & &(b)\\
\end{tabular}
\caption{
    Statistics of symmetric pairs $\{w,\bar{w}\}$
  in the 1\% top-ranked set of each divergence
	measure, for $k=7$: (a) average word pair
	frequency $(n^w + n^{\bar{w}})/2\;$ and
	(b)	frequency discrepancy $APR$.
	Complete genome.}
\label{fig:ranking_complete}
\end{figure}

Looking at the top-ranked sets for $k=7$ in
more detail shows specific differences.
In Fig.~\ref{fig:ranking_complete}(a) we see
that the 1\% top-ranked word pairs for $D_J$
and $D_E$ consist of words with low word
frequencies, whereas the 1\% top-ranked word
pairs for $D_P$ are composed of words with
much higher frequencies.
In Fig.~\ref{fig:ranking_complete}(b) we
note that the top-ranked word pairs for $D_P$
also have higher frequency discrepancy
values (absolute Pearson residuals).

A visual inspection of the distance distributions
in word pairs with high-ranked $D_J$ reveals
that there are many sparse distributions
among them. By sparse we mean that there are
many zero frequencies $f^w(i)$, and we already
saw that these words have a low total absolute
frequency.
Indeed, the dissimilarity measures $D_J$ and
$D_E$ may be overstating the disagreement
between distance distributions with local
differences. In fact, $D_J$ is quite sensitive
to small frequencies, while $D_E$ is sensitive
to the presence of a few high frequencies.
It should be noted that in the presence of
sparse distributions both low and high relative
frequency values are expected, which strongly
affect the results of $D_E$ and $D_J$.
On the other hand, $D_P$ ignores small
frequencies and evaluates the disagreement
between the sizes of the three strongest peaks,
which are taken into account even when their
locations do not precisely coincide.
Moreover, the peak size differences
are scaled by the highest peak sizes observed
in each distribution.

In view of these results, in what follows we
will focus on the dissimilarity measures $D_P$
and $APR$ for the detection of discrepancies
between symmetric word pairs.

\section{Detection of atypical symmetric
         word pairs}
In this section we focus on symmetric word
pairs consisting of words with length
$k$ = 5, 6, and 7, both in
the complete human genome
assembly and in a masked version.

In order to identify atypical words, we will
use three approaches.
First, we will consider the peak dissimilarity
between the distance distributions.
Second, we will combine this information with
the frequency discrepancy.
Finally, we will study the deviations between
the observed distance distributions and the distance
distributions under the assumption of randomness
and Chargaff's parity rule.

\subsection{Analyzing the observed peak
            dissimilarities}
As before, the peak dissimilarity is computed
with bandwidth $h=5$ and the $n=3$ strongest
peaks.
To capture the most dissimilar distance
distributions we select those symmetric word
pairs with peak dissimilarity above the
$99^{th}$ percentile of $D_P$ values.
This procedure captured 6 word pairs of length
$k=5$, 21 of length $k=6$ and 82 of length $k=7$.
Next, these words were sorted by decreasing
peak dissimilarity value. The results are listed
in Table~\ref{tab:list_complete} (for $k = 6$ and $k = 7$
only the first 20 results are shown).

\begin{table}[htbp]
\centering
\caption{Symmetric word pairs with peak
  dissimilarity above the $99^{th}$ percentile
	of $D_P$ values, by word length (only the
	first 20 results are shown). For each word
	$w$ its $D_P(w,\bar{w})$ value is given.
	Complete genome.}
\renewcommand{\arraystretch}{1.2}
\setlength\tabcolsep{2pt}
\scriptsize
\begin{tabular}{|lr|lrlr|lrlr|}
\hline
  k=5 & & k=6 & & & & k=7 & & & \\
\hline
  $w$ & $D_P$ & $w$ & $D_P$ & $w$ & $D_P$ &
	$w$ & $D_P$ & $w$ & $D_P$ \\
\hline
    CGAAG  & 127.9  & AGTATC & 91.0   & GAAATC & 58.7   & AAATTCC & 178.8  & AGGTTAA & 106.0 \\
    ACGAA  & 87.2   & AGTTAC & 86.4   & AAGGCC & 46.3   & ACTTTAC & 145.4  & AACAATC & 105.2 \\
    TACGA  & 43.5   & GGTTAA & 84.5   & CCTTCG & 46.3   & GCTTGAA & 138.9  & AAACTTA & 102.5 \\
    AACGG  & 37.0   & AGTAAC & 80.7   & ATACGA & 45.8   & CTGTCAA & 123.8  & GCAGTTA & 102.3 \\
    GAAAC  & 25.8   & GTTGGA & 80.6   & GTCACA & 45.1   & AACACAA & 120.4  & CTTGACA & 100.1 \\
    TCCAA  & 22.1   & ACCCGT & 69.1   & CTTCGA & 44.6   & AGTTTAA & 116.1  & GTAGAAC & 97.1 \\
           &        & AGGTTA & 68.2   & AAGTTA & 43.6   & GGGAAGA & 110.4  & AAATCCT & 96.8 \\
           &        & AAATCG & 65.9   & ACGAAG & 42.3   & GATGCCA & 107.7  & CGGGTTC & 96.3 \\
           &        & GAATAC & 61.2   & AGTCAC & 41.6   & CACTAAG & 107.5  & AAGGTTA & 95.0 \\
           &        & AGTCGA & 60.1   & CGGGTA & 39.4   & AACAGTA & 106.8  & ATTGGAG & 91.7 \\
\hline
\end{tabular}
\label{tab:list_complete}
\end{table}

\begin{figure}[htbp]
\centering
\begin{tabular}{cc}
  \includegraphics[width=5.7cm]
	  {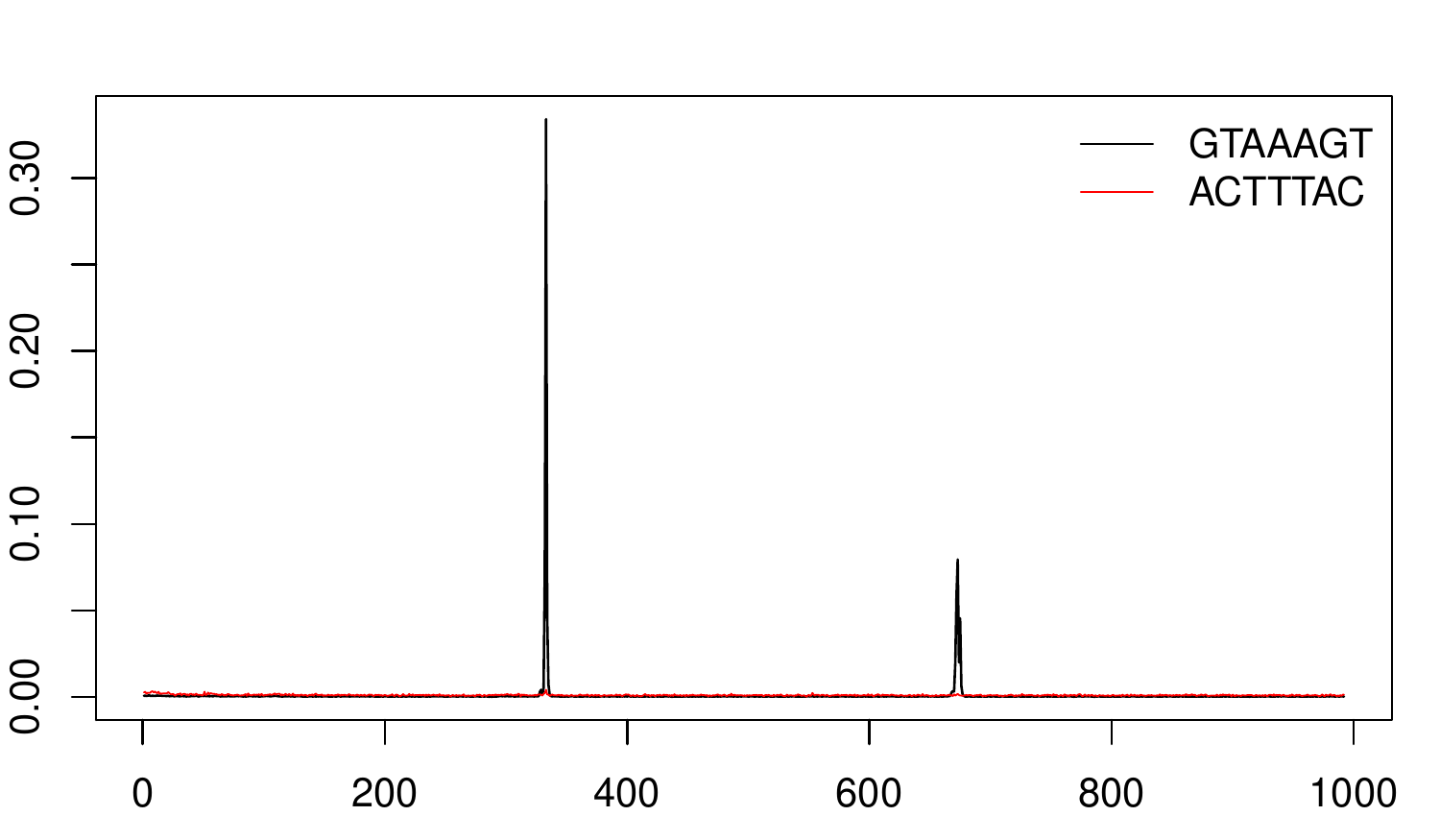} &
	\includegraphics[width=5.7cm]
	  {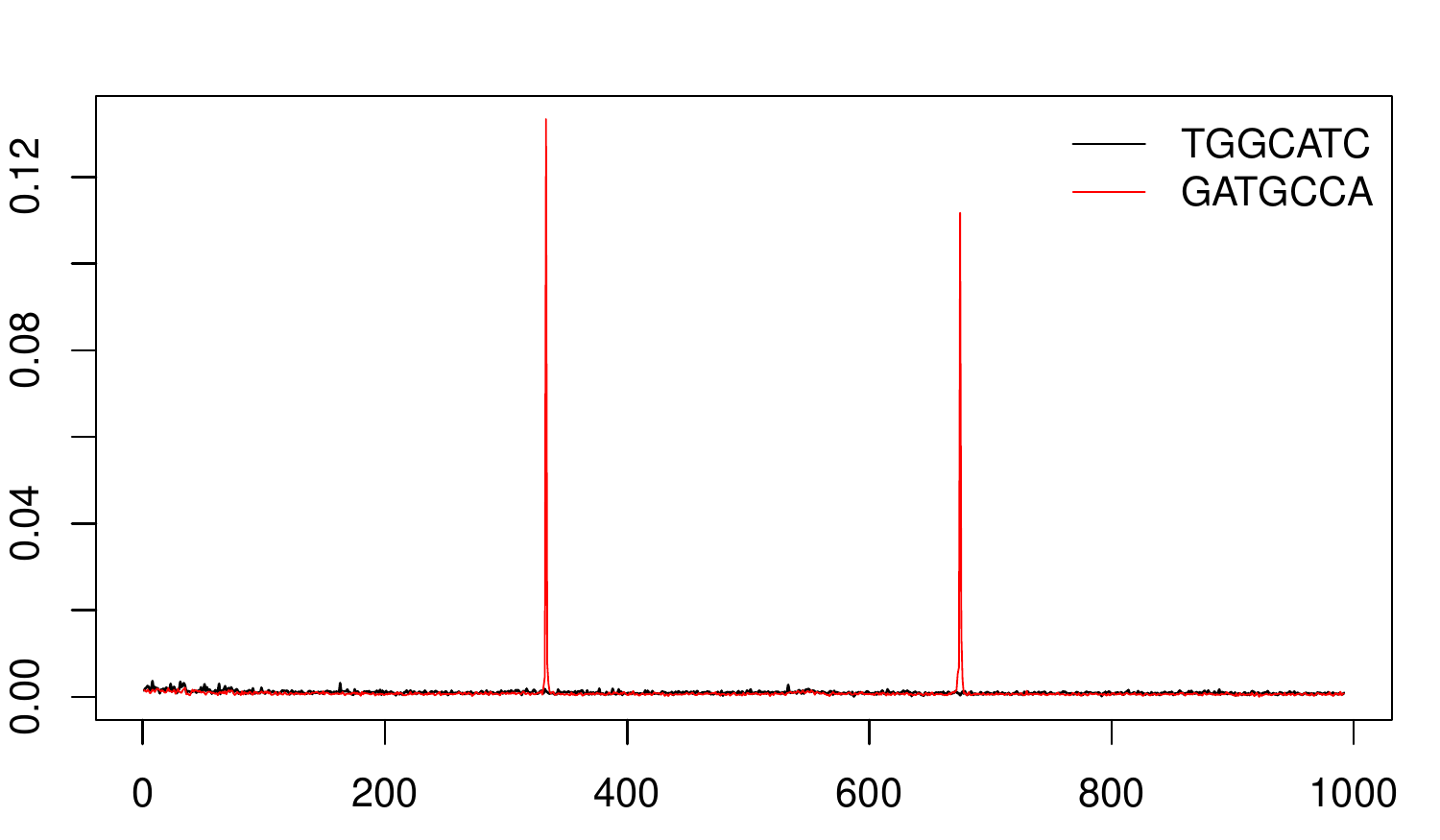}\\
	(a) &  (b)\\
  \includegraphics[width=5.7cm]
	  {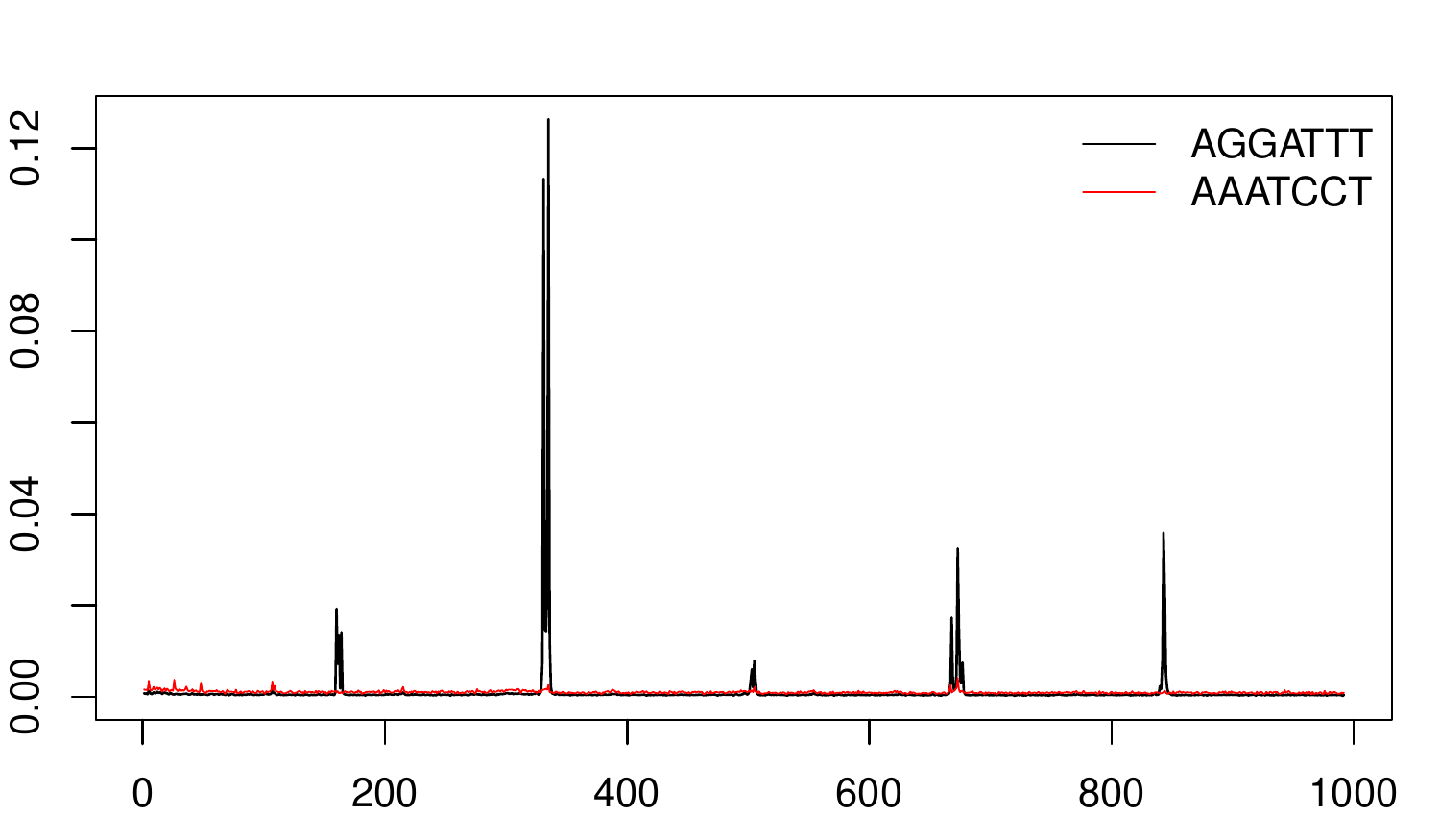}&
	\includegraphics[width=5.7cm]
	  {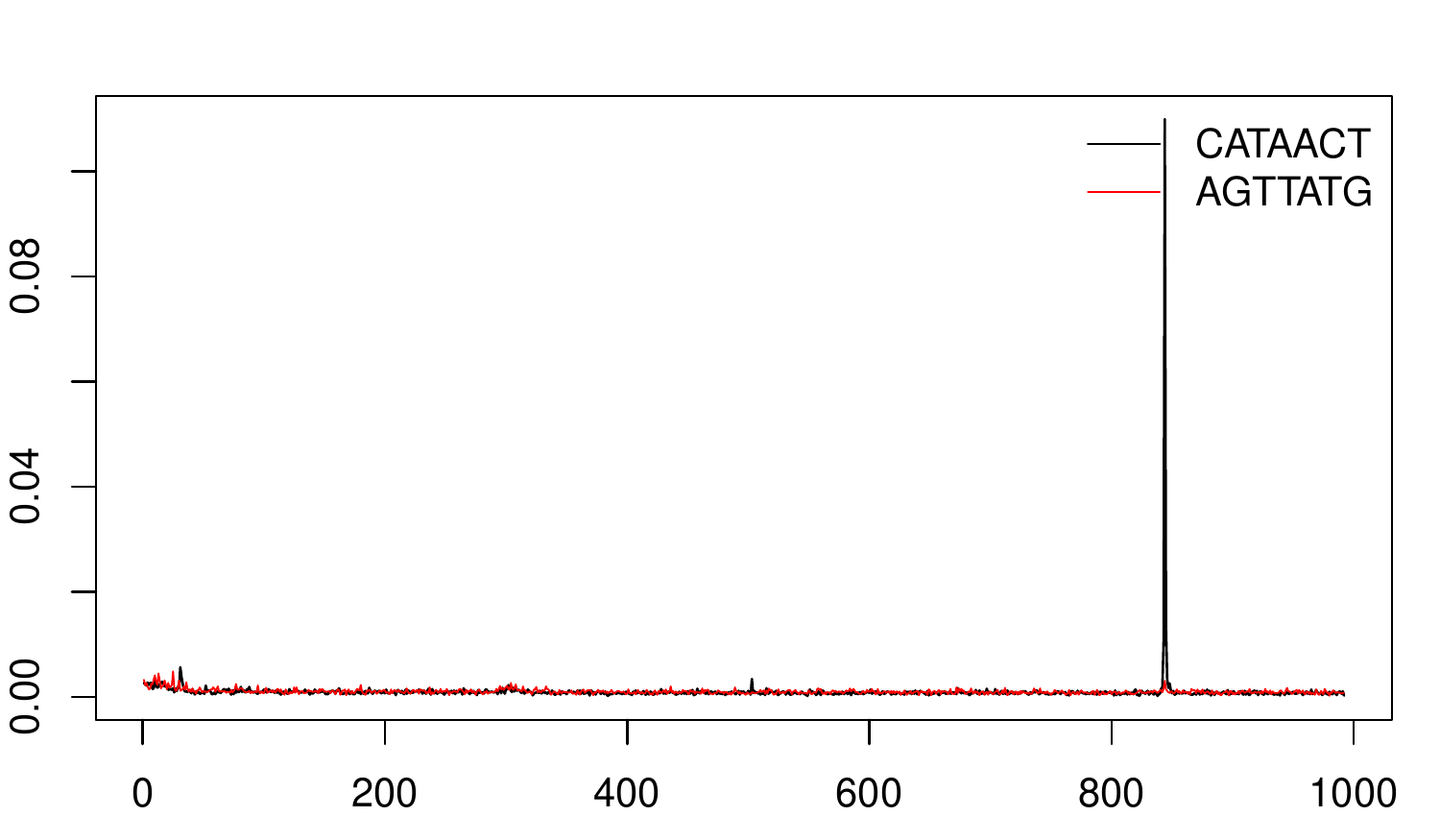}\\
	(c) &  (d)\\	
\end{tabular}
\caption{Distance distributions of some
  reverse complements, $f^w$ and $f^{\bar{w}}$,
	with high peak dissimilarity values:
	(a) $D_P=$145.4, $APR$=37.0; (b) $D_P=$107.6,
  $APR$=4.9; (c) $D_P=$96.8, $APR$=50.9;
	(d) $D_P=55.75$, $APR$=2.0. Complete genome.}
\label{fig:complete_dissimilar}
\end{figure}

Looking at these distributions, it turns out
that these high peak dissimilarities are
often caused by one distribution with strong
peak(s) and another displaying low variability
or small peaks, as illustrated
in Fig.~\ref{fig:complete_dissimilar}.

\begin{figure}[htbp]
\centering
\begin{tabular}{cc}
  \includegraphics[width=5.7cm]
	  {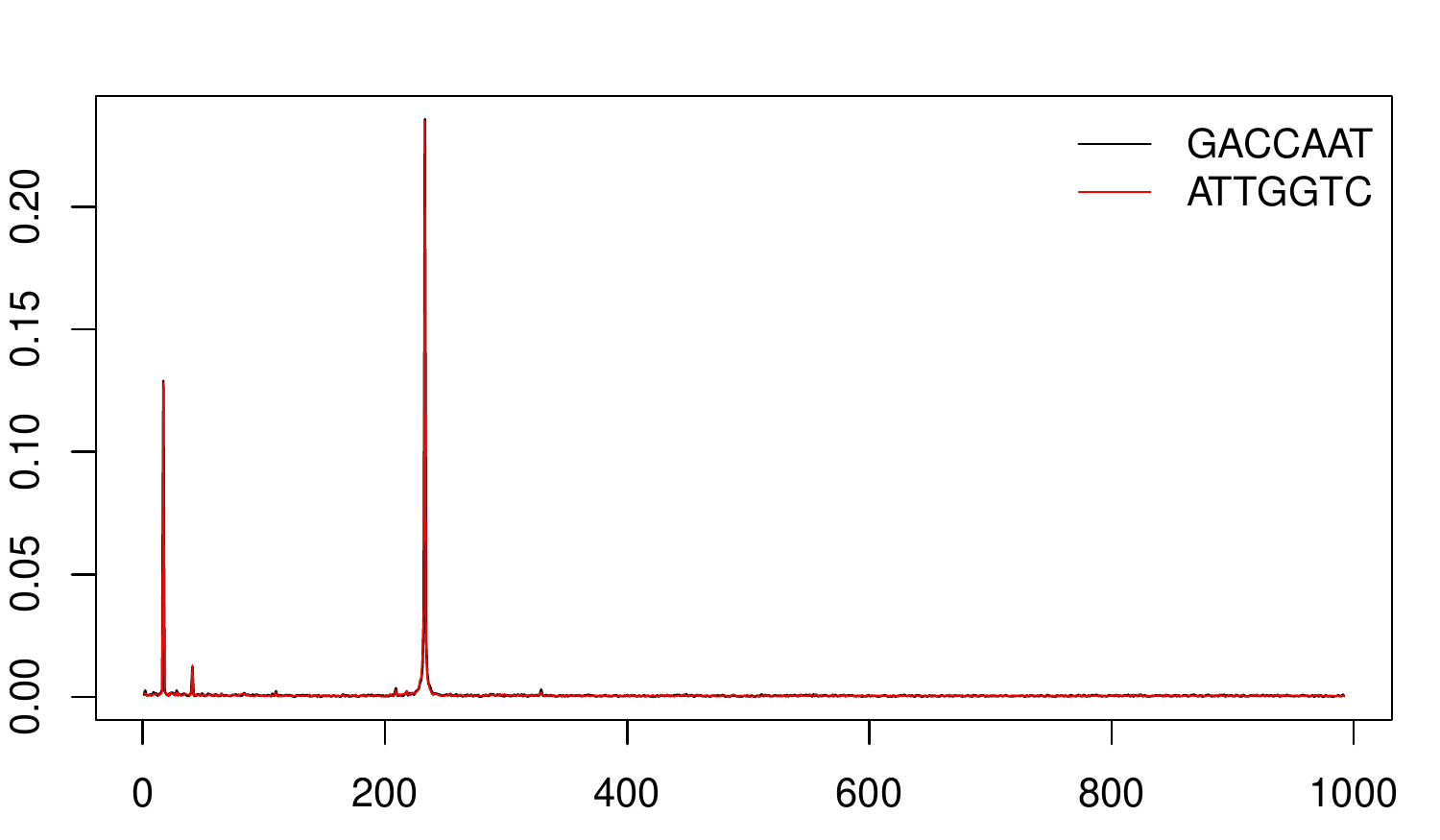} &
	\includegraphics[width=5.7cm]
	  {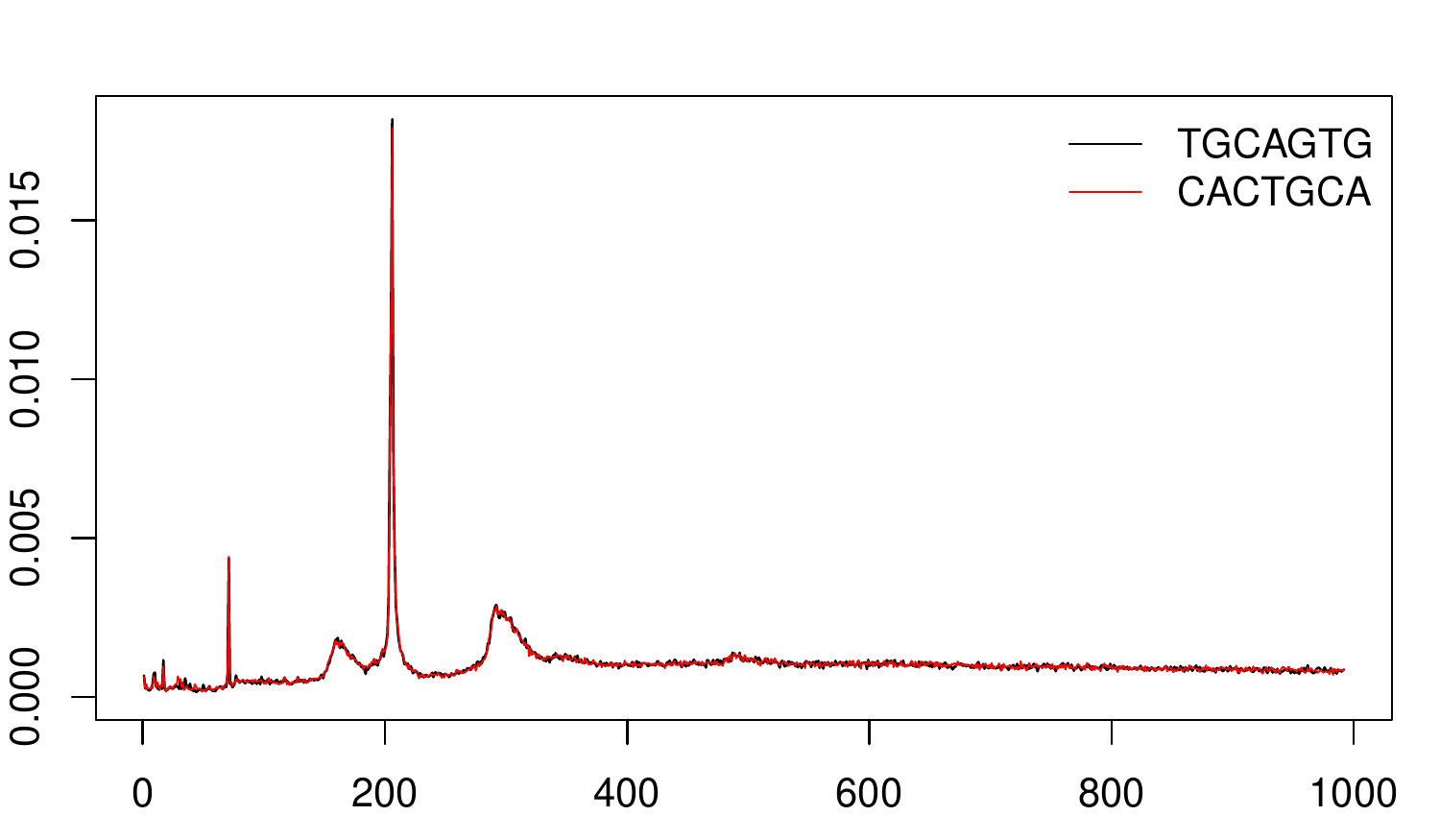}\\
	(a) &  (b)\\
  \includegraphics[width=5.7cm]
	  {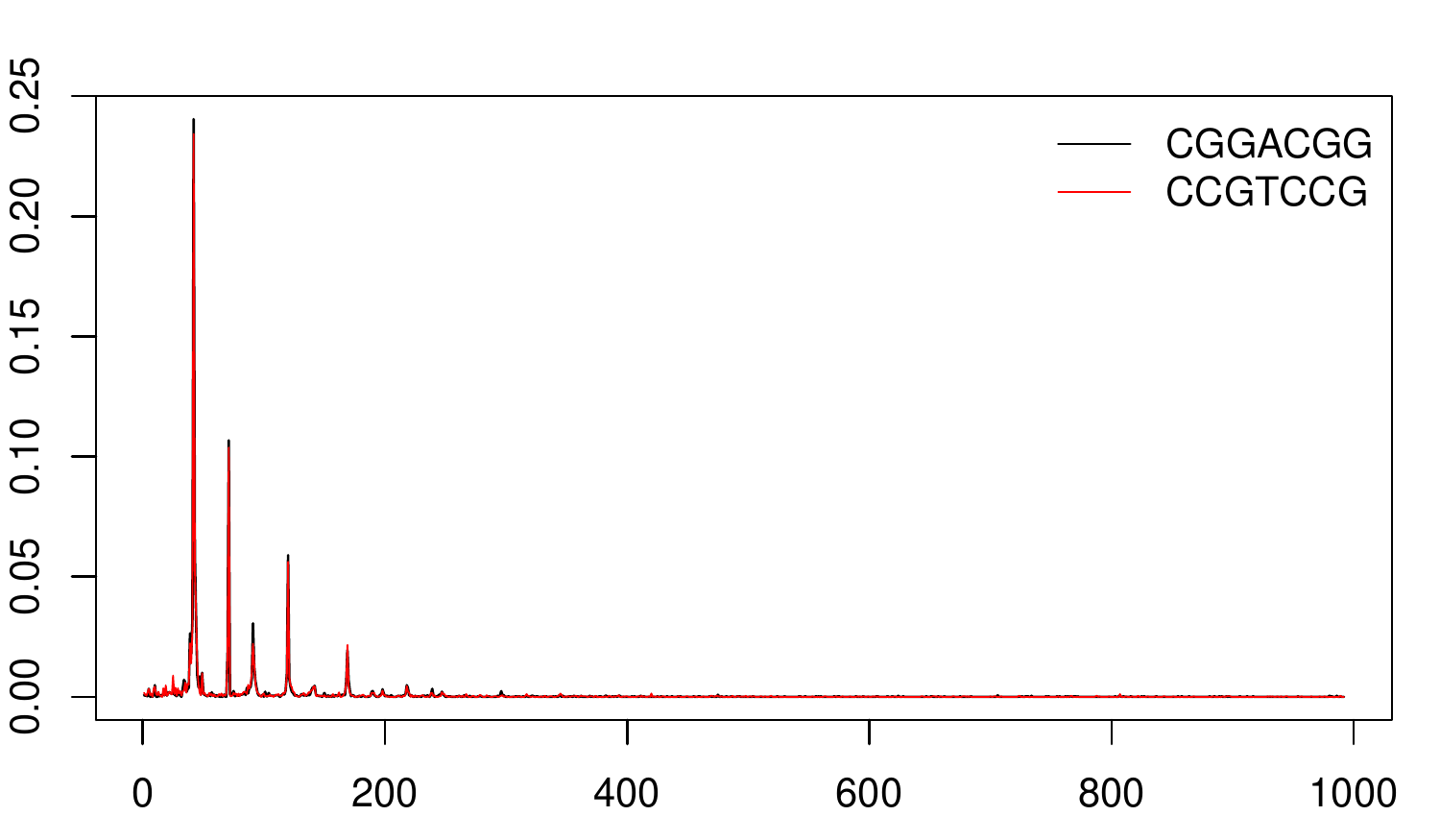} &
	\includegraphics[width=5.7cm]
	  {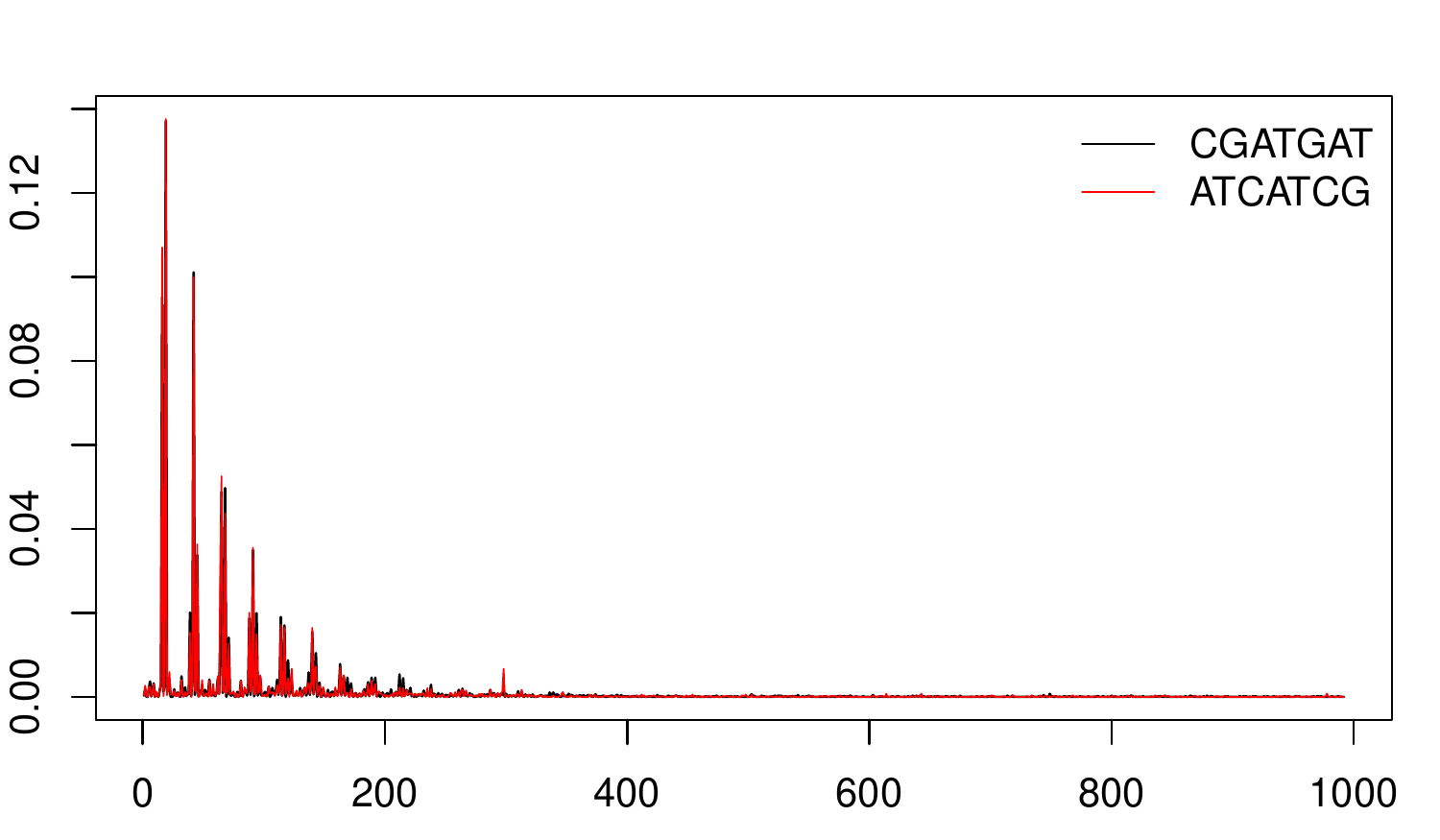}\\
	(c) &  (d)\\	
\end{tabular}
\caption{Distance distributions of some
  reverse complements, $f^w$ and $f^{\bar{w}}$,
	with low peak dissimilarity values:
	(a) $D_P$=0.012, $APR$=0.70;
	(b) $D_P$=0.026, $APR$=0.73;
	(c) $D_P$=0.060, $APR$=11.1;
	(d) $D_P$=0.116, $APR$=4.04.
	Complete genome.}
\label{fig:complete_similar}
\end{figure}

The symmetric pairs with low values of $D_p$ have very similar distributions.
For some words, this dissimilarity is surprisingly low in spite of their distance
distributions having irregular patterns and/or
some strong peaks.
Some of those distributions,
with peak	dissimilarities below the $10^{th}$
percentile of $D_P$, are illustrated
in Fig.~\ref{fig:complete_similar}.

\subsection{Combining peak dissimilarity
            and frequency discrepancy}
\begin{figure}[ht]
\centering
\includegraphics[width=0.32\textwidth,
  trim={0 0 0.5cm 0}, clip]
	{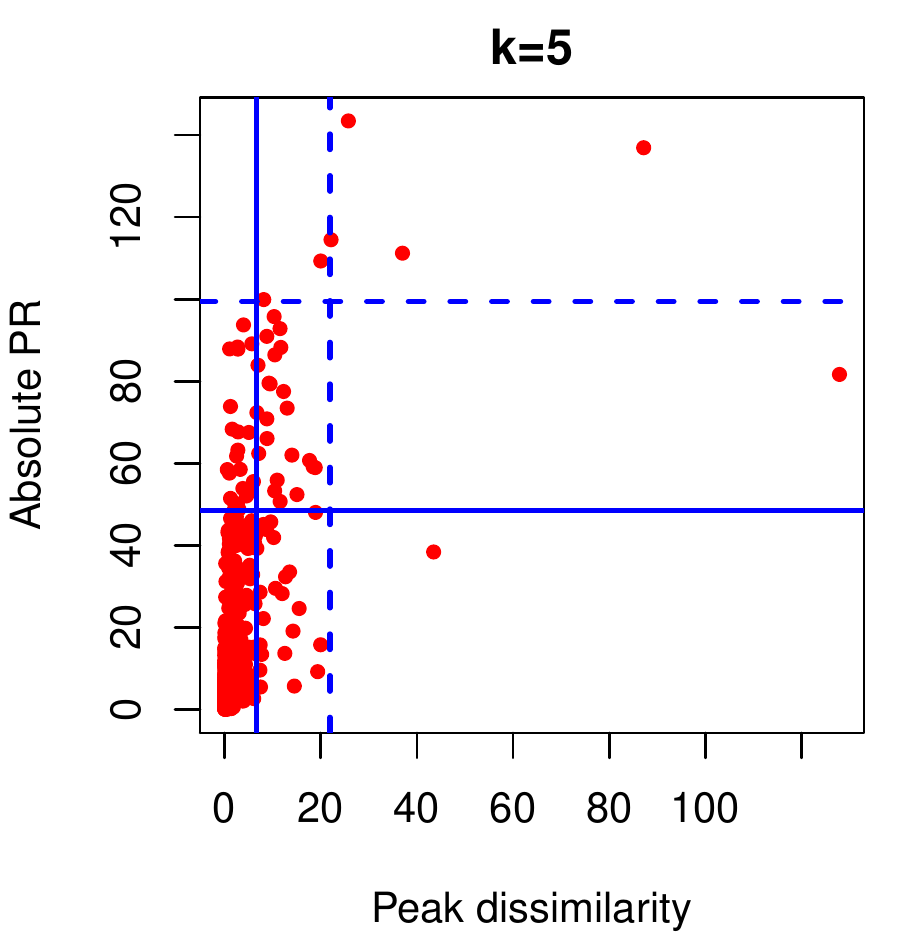}
\includegraphics[width=0.32\textwidth,
  trim={0 0 0.5cm 0}, clip]
  {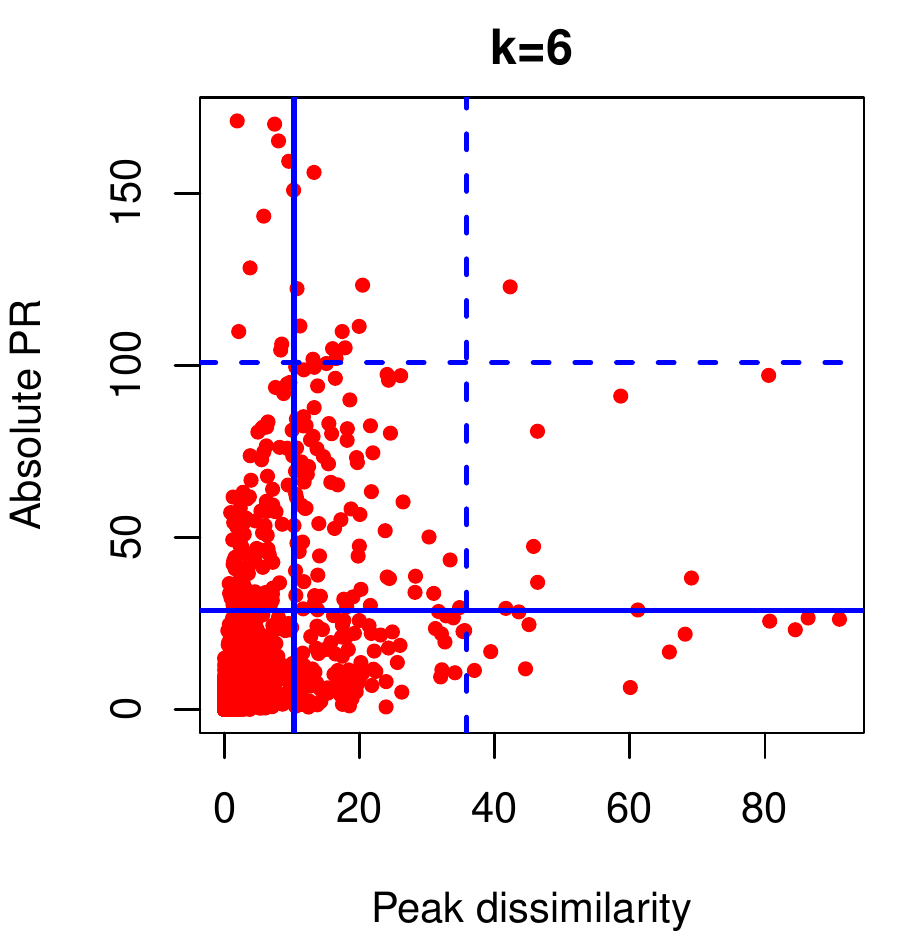}
\includegraphics[width=0.32\textwidth,
  trim={0 0 0.5cm 0}, clip]
  {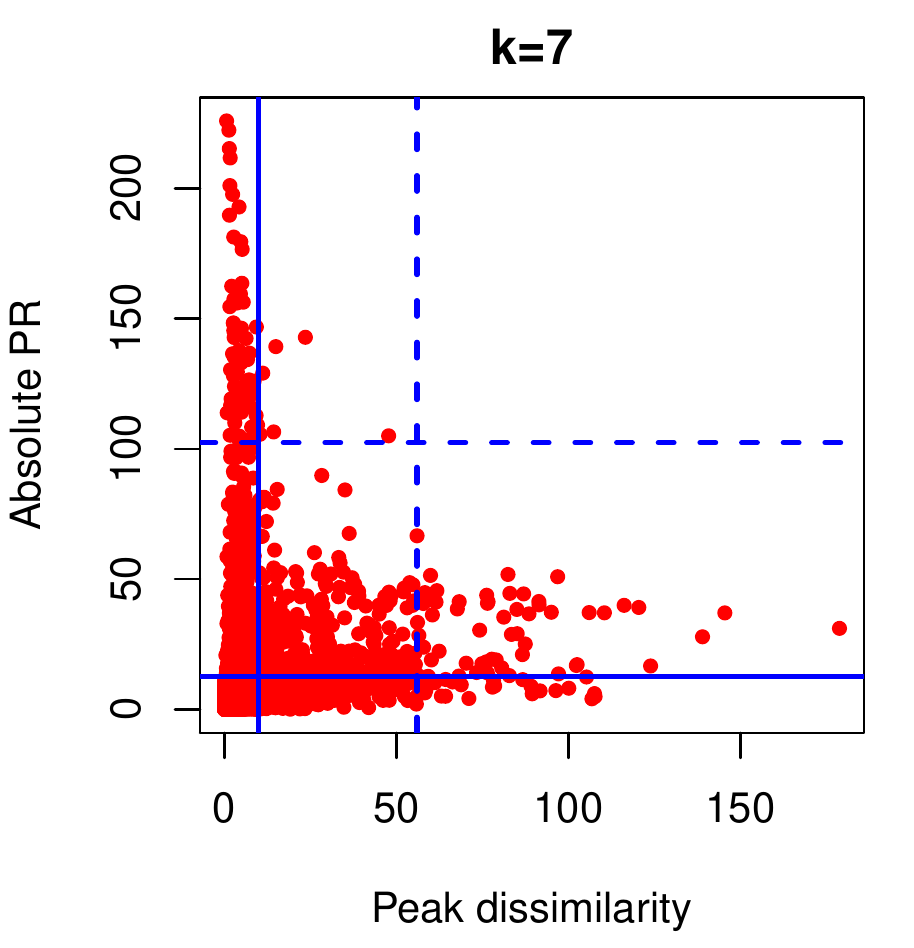}
\caption{Frequency discrepancy (APR) versus
  peak dissimilarity, for word lengths 5, 6
	and 7. Solid and dashed lines indicate the
	90$^{th}$ and the 99$^{th}$ percentile of
	each measure, respectively. Complete genome.}
\label{fig:scatter_skew_peak}
\end{figure}

In order to explore the (dis)similarity between
reverse complements we also combine the peak
dissimilarity $D_P$ with the frequency
discrepancy $APR$.
Fig.~\ref{fig:scatter_skew_peak} plots $D_P$
against $APR$ for each word length, with lines
indicating the $90^{th}$ and  $99^{th}$
percentile of both.
Whereas there is a kind of
positive relation between $D_P$ and $APR$ for
short words, this becomes less clear for
longer words, where we know that the rank
correlation between these measures decreases
(see Table~\ref{tab:spearman_complete}).

Several combinations of $APR$ and $D_P$ are
observed in
Fig.~\ref{fig:scatter_skew_peak}:
similar word
frequency with similar distance distribution
(call this case c1, which is common);
dissimilar word frequency with similar distance
distribution (c2); and similar word
frequency with dissimilar distance distribution
(c3).
(A fourth combination, dissimilar word
frequency and dissimilar distance distribution,
becomes increasingly rare for longer words.)

The interesting cases are (c2) and (c3),
which may reveal features of interest and
should be further studied.
In case (c2), words have similar
distance distributions but their frequencies of
occurrence are quite different, which
corresponds to points at the upper left
of Fig.~\ref{fig:scatter_skew_peak}.
To illustrate, consider the symmetric pair
with $w=CCGTCCG$
(Fig.~\ref{fig:complete_similar}.c), which
has peak dissimilarity below the $10^{th}$
percentile of $D_P$ and frequency
discrepancy around
the $90^{th}$ percentile of $APR$.
Conversely, in case (c3) strand symmetry
holds but the
words have distinct distance distributions along
the genome. This corresponds to points at
the bottom right of the plot.
For instance, the symmetric pair with
$w=AGTTATG$
(Fig.~\ref{fig:complete_dissimilar}.d) has
peak dissimilarity above the $90^{th}$
percentile of $D_P$ and frequency
discrepancy around the median of $APR$.
Observe that all word pairs listed in
Table~\ref{tab:list_complete} are located
on the right side of the scatter plot.

These results indicate that some asymmetries
in the human genome go far beyond Chargaff's
parity rule.

\subsection{Deviations from randomness}

It is intriguing that the distance distributions
of a symmetric pair can be very similar even
when their pattern is unexpected.
If genomic sequences were generated from
independent symbols only subject to
Chargaff's
parity rule ($\%A =\%T$ and $\%C =\%G$),
the inter-word distance distributions would be
close to an exponential distribution.
We are interested in investigating how dissimilar distance
distributions from such symmetric pairs
can be from the pattern under the random
scenario.
For that purpose, we compute the peak
dissimilarity between the averaged distance
distribution of the symmetric pair,
$(f^w+f^{\bar{w}})/2$, and the corresponding
averaged reference distribution.
The expected distance distribution can be deduced
using a state diagram, which represents the
progress made towards identifying $w$ as
each symbol is read from the sequence.
The input parameters are the nucleotide
frequencies in the sequence.
The algorithm used to construct those
reference distributions is a special case of
Fu's procedure based on finite Markov chain
embedding~\cite{fu1996}.

We select all symmetric pairs with
intra-pair peak dissimilarity below the
$10^{th}$ percentile of $D_P$, and ranked them
according to the peak dissimilarity between
their average distribution and their average reference
distribution (denoted as $rs$). This yields a list of symmetric
pairs with similar but unexpected distance
distributions. For each word length the top 20
results are listed in
Table~~\ref{tab:sim_unexpected}.
To illustrate some distance distribution of symmetric word pairs
with this behaviour, consider the pairs associated with
the words $w=CCGTCCG$
[Fig.~\ref{fig:complete_similar}(c)]
and $w=ATCATCG$ [Fig.~\ref{fig:complete_similar}(d)], which are
listed in this table under $k=7$.
The symmetric pairs have very similar distance distributions and their strong peaks make
them very dissimilar from the expected distributions in the random scenario.

\begin{table}[htbp]
\centering
\caption{Symmetric pairs with intra-pair peak
  dissimilarity below the $10^{th}$ percentile
	of $D_P$, sorted by decreasing dissimilarity
	to the random scenario (only the first 20
	results are shown)
	and organized by word length.
	For each word $w$ its $D_P(w,\bar{w})$ value
	is given and dissimilarity
	to the random scenario ($rs$). Complete genome.}
\renewcommand{\arraystretch}{1.2}
\setlength\tabcolsep{2pt}
\scriptsize
\begin{tabular}{|lll|lll|lll|}
\hline
    k=5 & & & k=6 & & & k=7 & & \\
\hline
    $w$    & $D_P$  & $rs$    & $w$    & $D_P$& $rs$    & $w$     & $D_P$ & $rs$ \\
    CGCCC  & 0.009   & 213.80 & CGCCCG & 0.029 & 583.44 & ACGCGTA & 0.141  & 1621.58\\
    CCTCC  & 0.015   & 207.89 & CGGGAG & 0.018 & 443.79 & CAACGAG & 0.122  & 1556.41\\
    CGGCC  & 0.014   & 206.40 & GCCTCC & 0.005 & 418.84 & CTCGAGA & 0.160  & 1481.80\\
    CCAGC  & 0.009   & 190.02 & AGGCCG & 0.014 & 360.64 & ATCGCCA & 0.082  & 1350.15\\
    CCTCG  & 0.025   & 184.80 & CAGACG & 0.012 & 354.04 & CGTCTGA & 0.130  & 1292.38\\
    CGCCA  & 0.014   & 174.63 & CAGGAG & 0.012 & 339.94 & ACGCAAA & 0.056  & 1257.21\\
    CCGCC  & 0.014   & 153.47 & GGTCTA & 0.034 & 332.90 & GTTCGGA & 0.120  & 1097.62\\
    CAGGC  & 0.008   & 136.91 & AGATCG & 0.024 & 326.56 & \underline{ATCATCG} & 0.116 & 1040.96\\ 
    GCCGA  & 0.024   & 136.10 & CGAGAC & 0.025 & 291.41 & CATCGAA & 0.111  & 1038.82\\
    CCCGG  & 0.021   & 133.17 & CACGCC & 0.038 & 289.29 & TCATCGA & 0.143  & 1031.44\\
    CCACC  & 0.023   & 115.13 & CCCGTC & 0.037 & 276.62 & AGGAGCG & 0.099  & 995.72\\
    CTCCC  & 0.018   & 103.37 & ACGGGG & 0.041 & 267.93 & CAGACGA & 0.120  & 957.98\\
    CCCAG  & 0.011   & 95.68  & CGTCTC & 0.009 & 266.46 & TCCCGGA & 0.025  & 904.82\\
    AGGAG  & 0.011   & 88.48  & GAGGCA & 0.018 & 265.75 & GGATCTA & 0.138  & 893.08\\
    GGCCA  & 0.014   & 87.62  & CCTCCC & 0.015 & 260.13 & CCGGACG & 0.099  & 892.40\\
    CAGGA  & 0.013   & 83.81  & CTCGGC & 0.021 & 258.12 & ACGCTCC & 0.096  & 891.33\\
    CCGAG  & 0.024   & 78.98  & CCCGGC & 0.030 & 246.31 & AGACGCT & 0.064  & 886.83\\
    CCAGG  & 0.027   & 74.37  & CCGGGC & 0.029 & 242.70 & \underline{CCGTCCG} & 0.060 & 866.16\\ 
    CTGCC  & 0.021   & 66.48  & CCCGGA & 0.042 & 242.56 & CAGACGG & 0.009  & 855.86\\
    AGTAG  & 0.005   & 64.42  & CGCCTC & 0.034 & 231.77 & CGGGCGC & 0.030  & 840.74\\  \hline
\end{tabular}%
\label{tab:sim_unexpected}%
\end{table}%

\subsection{Masked Genome Assembly}
To reduce the effect of repetitive sequences in the
original genome assembly, we also analyze a masked
version
of the genome which excludes major known classes of
repeats~\cite{lander2001}, such as long and short
interspersed nuclear elements (LINE and SINE), long
terminal repeat elements (LTR), Satellite repeats or
Simple repeats (micro-satellites). All distributions
and measures in this subsection are from the
masked sequence and for $k=7$.

\begin{figure}[htbp]
\centering
\begin{tabular}{ccc}
\includegraphics[width=0.3\textwidth,
trim={0 0 0.3cm 0},clip]
 {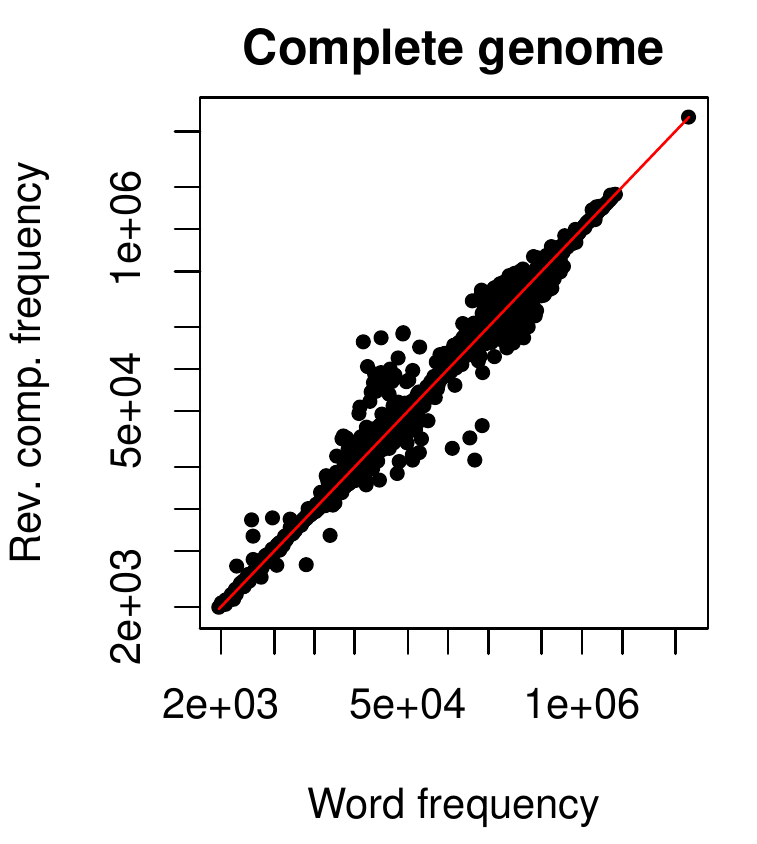} &
\includegraphics[width=0.3\textwidth,
trim={0 0 0.3cm 0},clip]
 {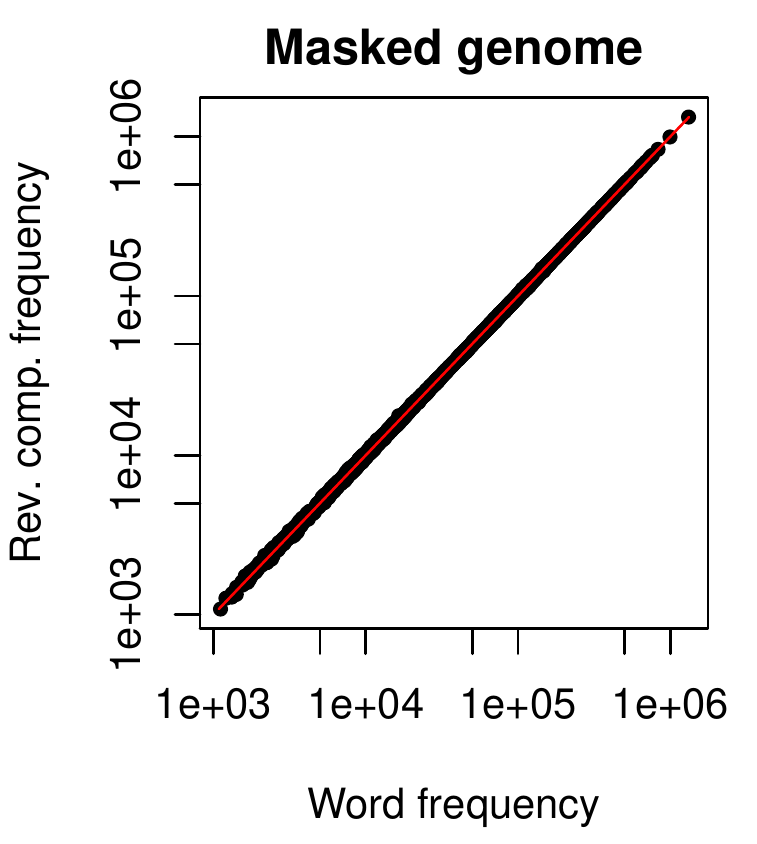} &
\includegraphics[width=0.3\textwidth,
trim={0 0 0.3cm 0},clip]
 {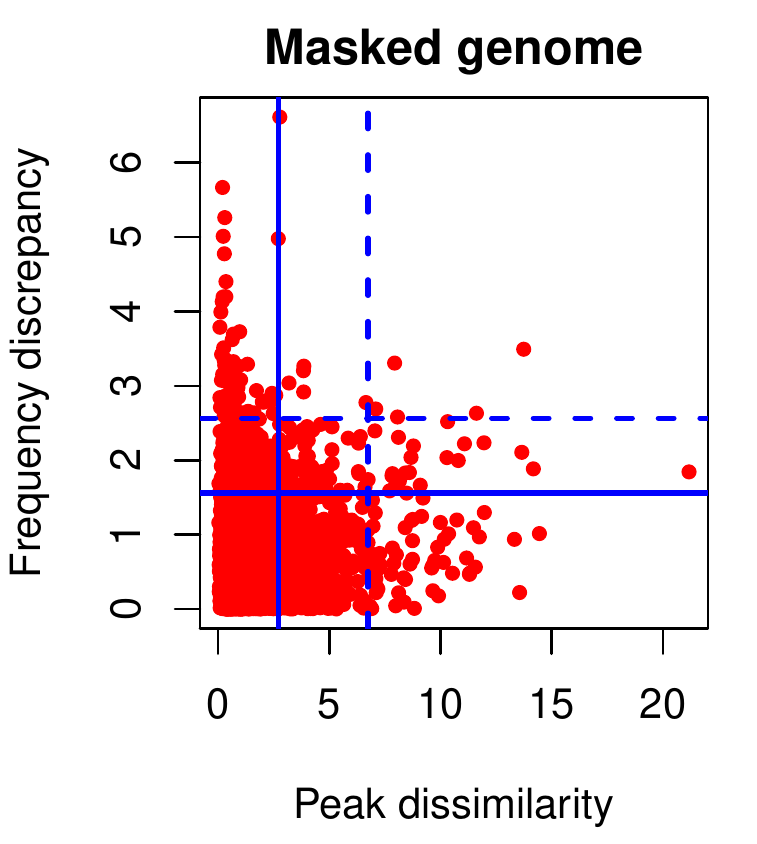}\\
 (a) & (b) & (c) \\
\end{tabular}
\caption{
(a) Word frequencies ($n^w$) in the
entire genome against those observed for the
reverse complements ($n^{\bar{w}}$) with both
axis in log scale, all for $k=7$; (b) Same for
the masked genome; (c) Frequency discrepancy
versus peak dissimilarity for $k=7$ in the
masked genome, where solid lines indicate the
90$^{th}$ percentile of each quantity.}
\label{fig:scatter_line_mask_all}
\end{figure}

Masking the genome sequence markedly affects the
shape of the distance distributions. Several strong peaks
observed in the complete genome are eliminated by
masking, as described in \cite{tavares2017pacbb}.
It also greatly reduces the frequency discrepancy
between reverse complements.
To visually inspect those discrepancies, we plot
the word frequencies against those observed for the
reverse complement. We observe that, for the
masked genome, the points are located much closer
to the diagonal line than in the complete genome
[Fig.~\ref{fig:scatter_line_mask_all} (a) and (b)].

\begin{figure}[htbp]
\centering
\begin{tabular}{cc}
  \includegraphics[width=5.7cm]
   {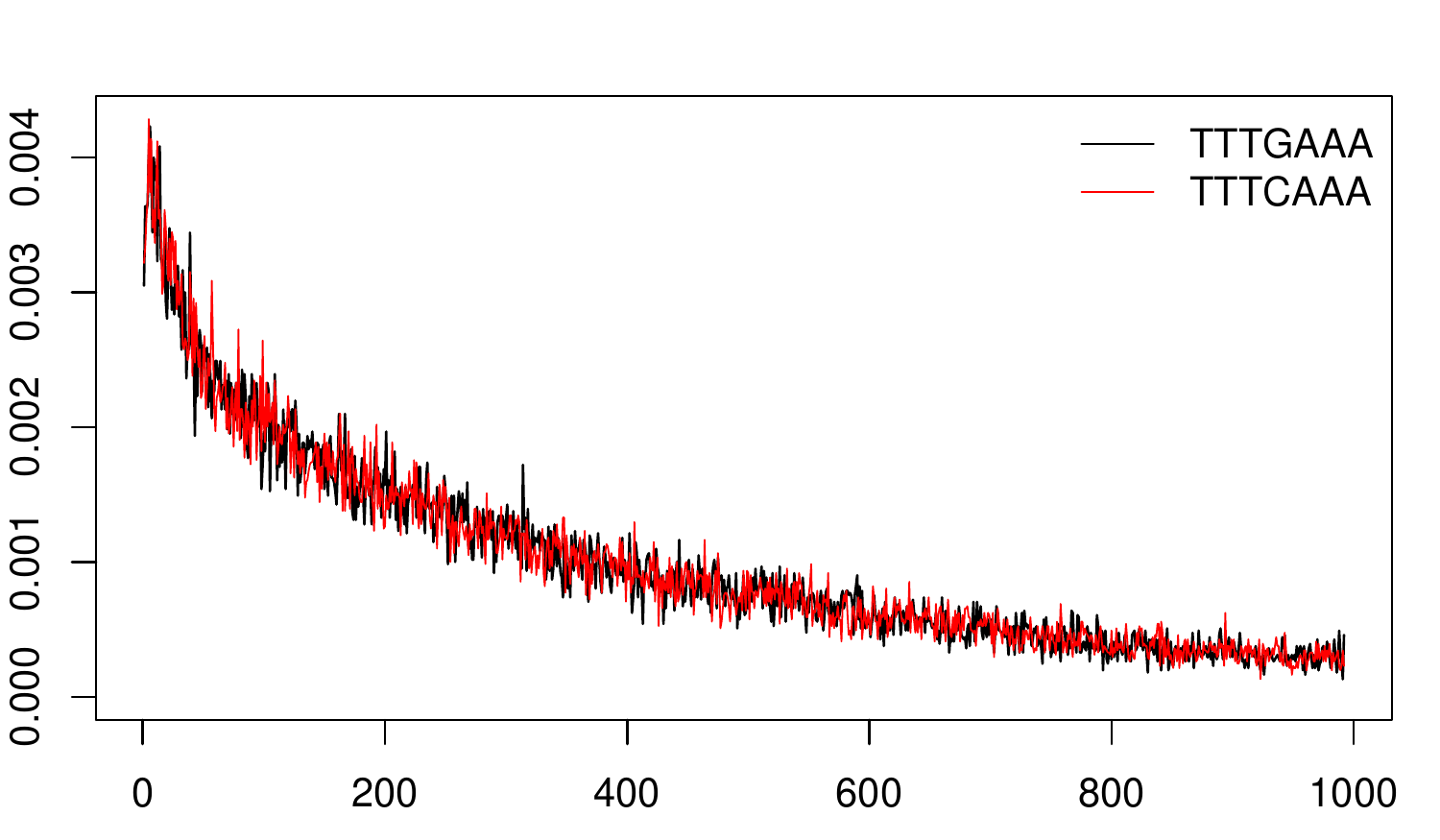} &
	\includegraphics[width=5.7cm]
   {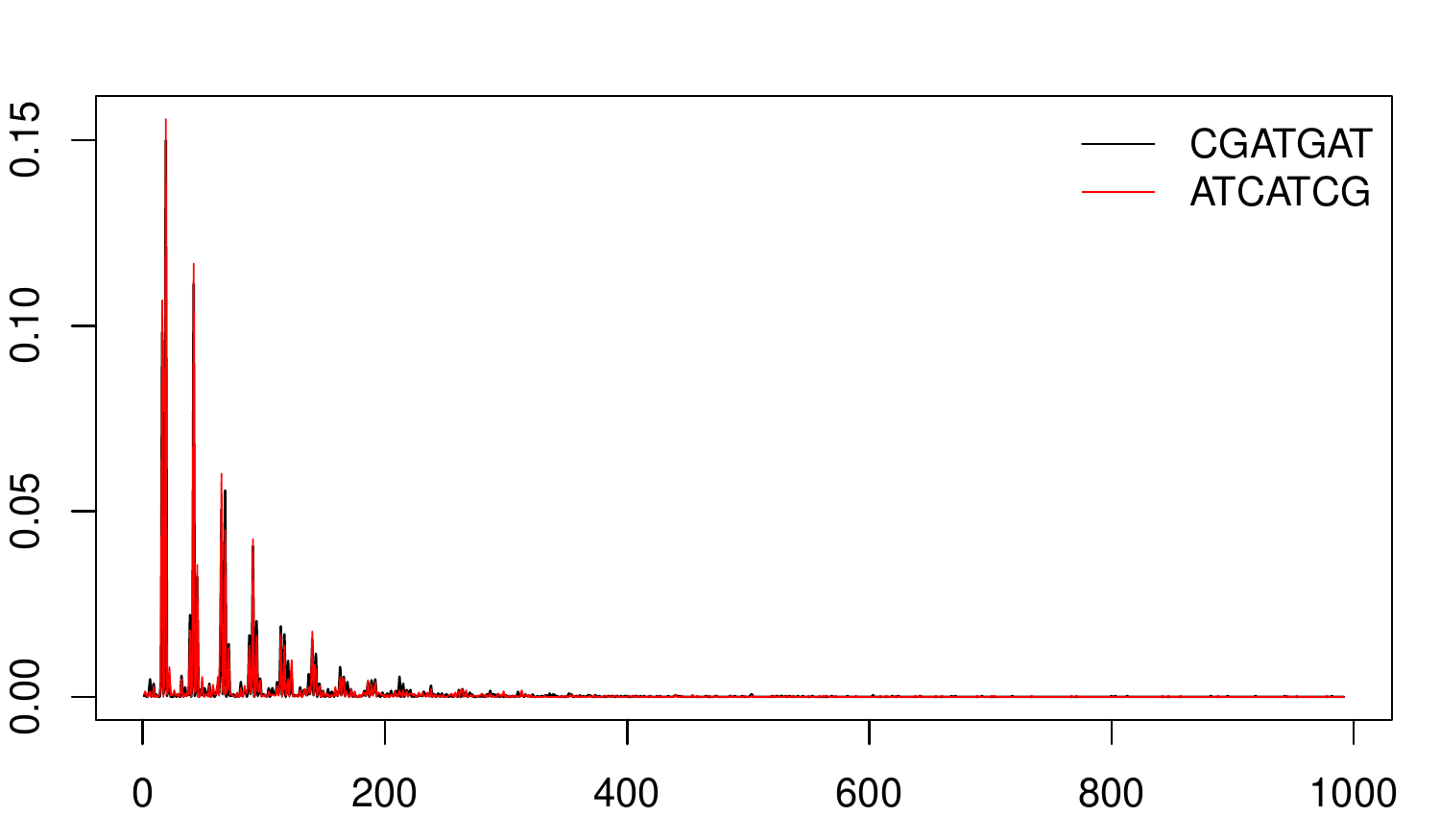}\\
	(a) &  (b)\\
  \includegraphics[width=5.7cm]
   {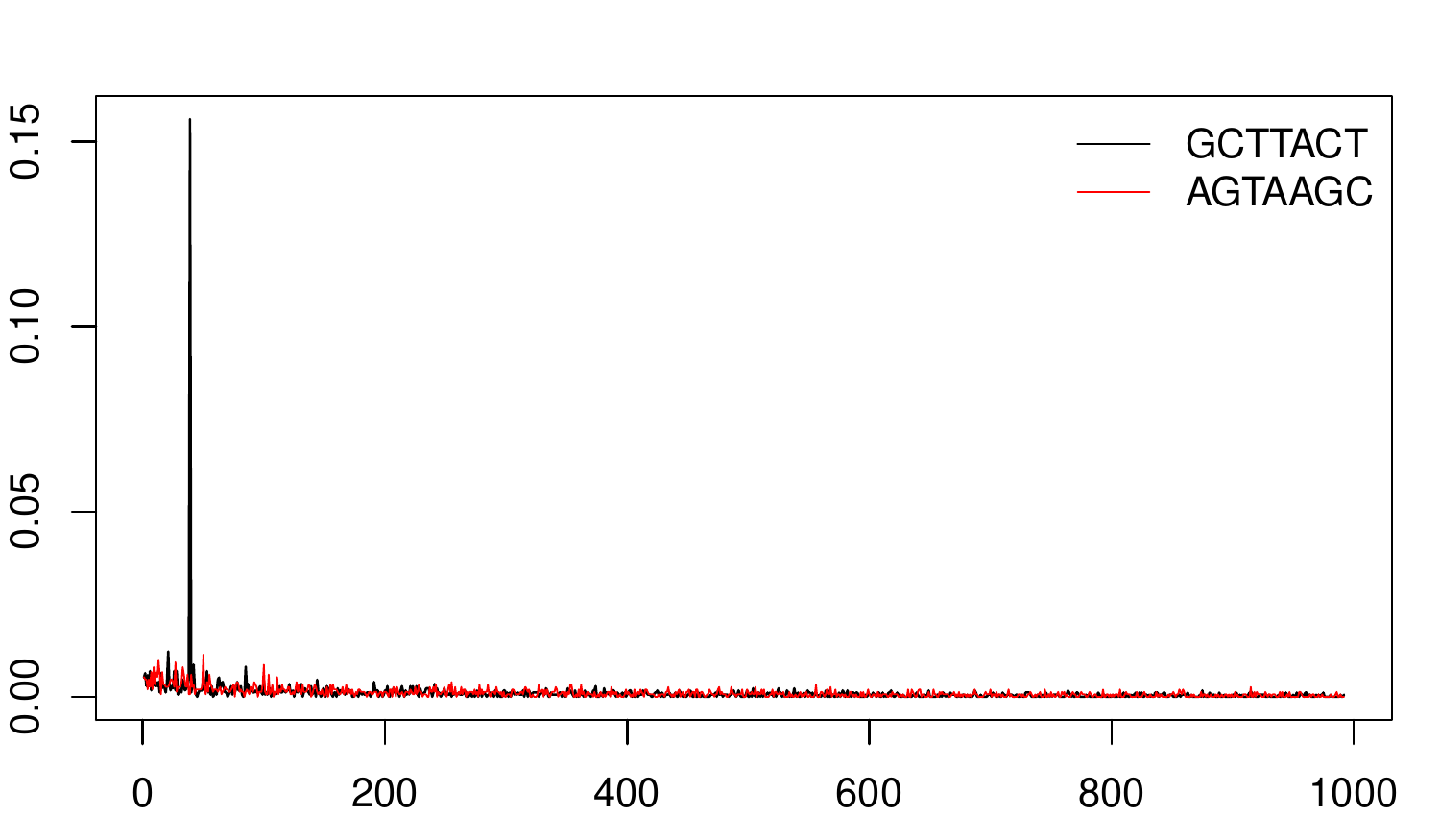} &
	\includegraphics[width=5.7cm]
   {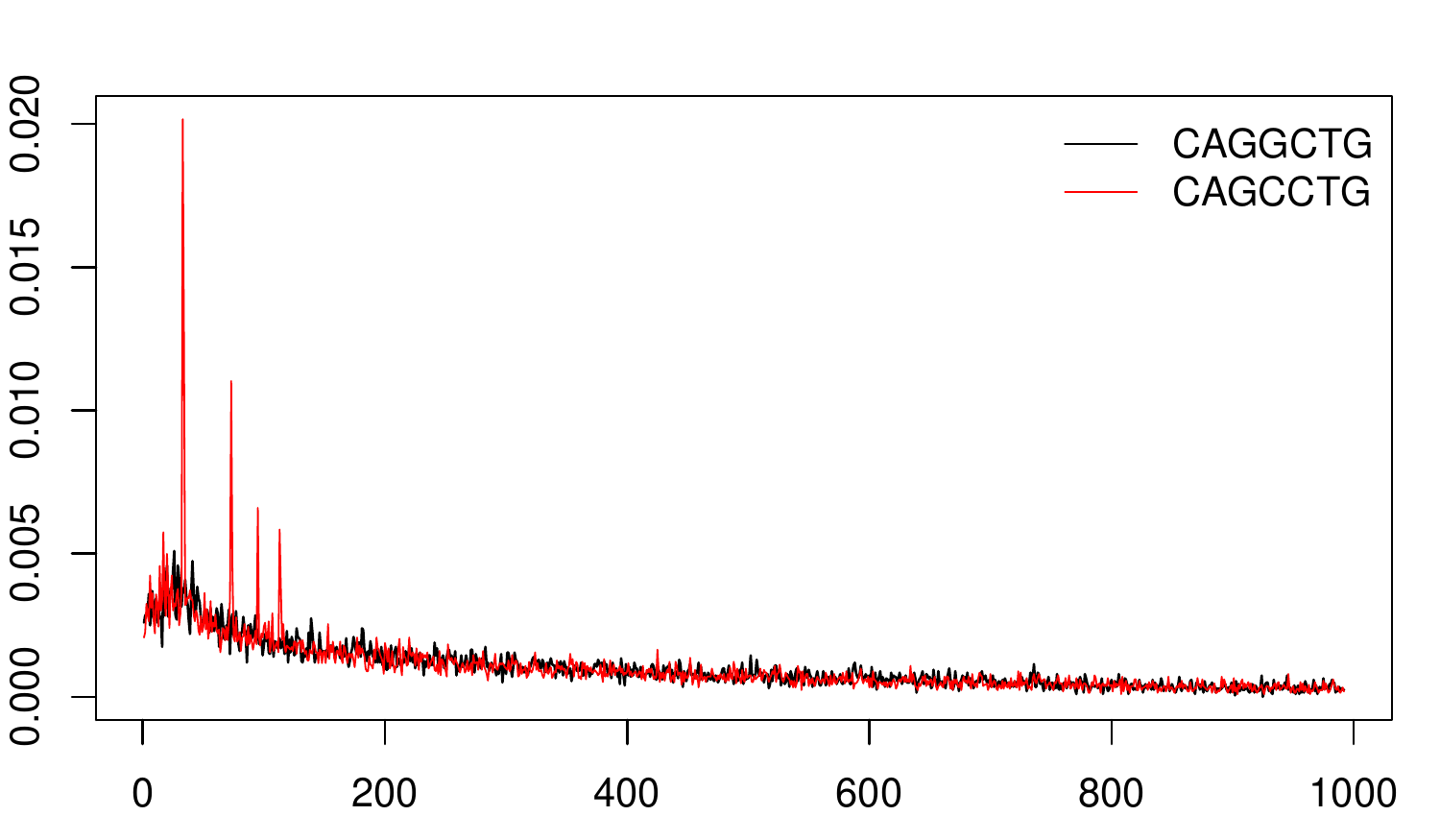}\\
	(c) &  (d)\\	
\end{tabular}
\caption{Distance distributions of some reverse
complements with low dissimilarity values:
0.144~(a), 0.125~(b); and with high
dissimilarity values: 11.74~(c), 6.49~(d).
Masked genome.}
\label{fig:mask}
\end{figure}

To select symmetric pairs with similar
and dissimilar distance distributions the authors in
\cite{tavares2017pacbb}
retained word pairs with peak dissimilarity
below the $10^{th}$ percentile of $D_P$
values and those above the $90^{th}$
percentile of $D_P$ values, after filtering
out words with low total absolute frequency.
They distinguish between two groups of word
pairs with low peak dissimilarity: those
where both distributions have strong peaks
at short distances, and on those where neither
distribution has strong peaks.
These patterns are illustrated in
Fig.~\ref{fig:mask}(a--b).
Interestingly, the unusual pattern of
$w=ATCATCG$ in the complete sequence
[Fig.~\ref{fig:complete_similar}(d)]
remains in the masked sequence
[Fig.~\ref{fig:mask}(b)].
Symmetric pairs with high dissimilarity
usually have one distribution with one or
more strong peaks at short distances ($<200$)
whereas the other presents low variability.
Some very dissimilar pairs are shown in
Fig.~\ref{fig:mask}(c--d).

\subsubsection{Annotation Analysis}
To investigate whether an association exists
between dissimilar reverse complements and
functional DNA elements, we perform an
annotation analysis for the 15 most dissimilar
symmetric pairs.
For each such pair we list the word with the
strongest peaks. Then we look for the
`favored' distance(s), i.e. those where the
strongest peak(s) are located.
These peaks are often concentrated in one
chromosome rather than being spread over the
entire genome sequence.
Table~\ref{tab:bio} lists the chromosome in
which the favored distances are most pronounced,
for each of the 15 pairs. The positions of
the words occurring at that distance from each
other are recorded.
Then, we retrieve annotations within these
genomic coordinates from UCSC GENCODE v24.
Interestingly, the words we obtain that
are located on chromosome 13 all fall within
the gene LINC01043 (long intergenic
non-protein coding RNA 1043) and all of our
words on chromosome 1 are contained in the
gene TTC34 (tetratricopeptide repeat domain
34). These results suggest that the most
dissimilar distributions may be related to
repetitive regions associated with RNA or
protein structure.

\begin{table}[ht]
\vspace{-0.15cm}
\centering
\caption{The 15 most dissimilar
 symmetric pairs with $k=7$, characterized
 by their word with the strongest peaks.
 The chromosome on which these peaks are
 prominent is indicated. Masked sequence.}
\renewcommand{\arraystretch}{1.2}
\setlength\tabcolsep{2.5pt}
\scriptsize
\begin{tabular}{|l|cc|c|c|c|c|}
\hline
 chromosome & \multicolumn{2}{|c|}{13} & 1
      & 4      & 3      & 8 \\
\hline
   word $w$ & $ACCATTC\:$ & $GGTAAGC\:$ &
	$AGCATCT\:$ & $GTTGGTA\:$
	& $TGGTATG\:$ & $GCTTACT\:$ \\
  & $CTTCAGG\:$ & $TAAGCAT\:$ & $GAGCATC\:$
	& $TGGTAGA\:$ & & \\
  & $GACCATT\:$ & $TCAGGAT\:$ & $TGAGCAT\:$
	&        &  & \\
	& $TCCTTCA\:$ & $TTCAGGA\:$ & & & & \\
\hline
\end{tabular}
\label{tab:bio}
\vspace{-0.15cm}
\end{table}

A deeper investigation into the biological
meaning of these words is necessary to
investigate whether the observed
dissimilarities reflect the selective
evolutionary process of the DNA sequence.

\section{Conclusions}
In this work we explore the DNA symmetry
phenomenon in the human genome, by comparing
each inter-word distance distribution to the
distance distribution of its reverse
complement, for word lengths $k=5$, 6 and 7.

We use the peak dissimilarity to evaluate
the dissimilarity between the distance
distributions of reverse complements and
compare it to two well-known measures.
Our results suggest that peak dissimilarity
achieves its intended purpose in the
detection of highly dissimilar distance
distributions.

In the complete human genome, we confirm
the existence of symmetric word pairs with
quite distinct distance distributions. In such
cases, one of the distance distributions
typically has well defined peaks and the
other has low variability. We also report
symmetric pairs with very similar distance
distributions even though these
distributions are themselves unexpected
with strong peaks.

The association between distance distribution
dissimilarity and frequency discrepancy is
analyzed.
In general, the correlation between those
measures is moderate. Several behaviors
are observed in symmetric pairs, by
combining low and high values of both
measures.
In particular there are symmetric pairs
that preserve strand symmetry (similar
frequency) but have dissimilar distance
distributions; and symmetric pairs with
dissimilar frequencies and similar distance
distributions. Symmetric pairs with
either behavior may uncover features of
interest.

We also investigate how well our results
hold up in a masked sequence, which
excludes major known classes of repeats.
Even though masking generally reduces the
dissimilarity between distance distributions
of symmetric pairs, there remain quite a
few word pairs with high dissimilarity,
which in our study are mainly localized
on a specific chromosome and even a
specific gene. A question worth
investigating is to what extent the high
dissimilarities may be linked to
evolutionary processes.

Taken together, our results suggest that
some asymmetries in the human genome go
far beyond Chargaff’s rules. Of particular
note are some symmetric pairs with a
perfectly ordinary frequency similarity
and distribution similarity, that exhibit
a strong preference for occurring at
some particular distances.


\section{acknowledgements}
This work was partially supported by the
Portuguese Foundation for Science and
Technology (FCT), Center for Research \&
Development in Mathematics and
Applications (CIDMA),
Institute of Biomedicine (iBiMED) and
Institute of Electronics and Telematics
Engineering of Aveiro (IEETA), within
projects UID/MAT/04106/2013,
UID/BIM/04501/2013 and
UID/CEC/00127/2013.
A. Tavares acknowledges the PhD grant
PD/BD/105729/2014 from the FCT.
The research of P. Brito was financed by
the ERDF - European Regional Development
Fund through the Operational Programme
for Competitiveness and
Internationalization - COMPETE 2020
Programme within project
POCI-01-0145-FEDER-006961, and by the FCT
as part of project UID/EEA/50014/2013.
The research of J. Raymaekers and
P. J. Rousseeuw was supported by projects
of Internal Funds KU Leuven.

\end{document}